\documentclass{emulateapj}
\usepackage{amsmath}
\usepackage{graphicx}
\begin{document}

\title{Protostellar Disk Evolution Over Million-Year Timescales
with a Prescription for Magnetized Turbulence}

\author{Russell Landry}
\affil{Physics Department, University of Texas at Dallas}
\email{russell.landry@gmail.com}
 
\author{Sarah E. Dodson-Robinson}
\affil{Astronomy Department, University of Texas at Austin}

\author{Neal J. Turner}
\affil{Jet Propulsion Laboratory/California Institute of Technology}

\and

\author{Greg Abram}
\affil{Texas Advanced Computing Center, University of Texas at
Austin}

\begin{abstract}
Magnetorotational instability (MRI) is the most promising
mechanism behind accretion in low-mass
protostellar disks. Here we present the first analysis of the global
structure and evolution of non-ideal MRI-driven T-Tauri disks on
million-year timescales. We accomplish this in a 1+1D simulation
by calculating magnetic diffusivities and utilizing turbulence
activity criteria to determine 
thermal structure and accretion rate without resorting to a 3-D magnetohydrodynamical (MHD) simulation.
Our major findings are as follows. First, even for
modest surface densities of just a few times the minimum-mass solar
nebula, the dead zone encompasses the giant planet-forming region,
preserving any compositional gradients. Second, the surface density of the active layer is nearly constant in
time at roughly 10~g~cm$^{-2}$, which we use to derive a simple
prescription for viscous heating in MRI-active disks for those who wish
to avoid detailed MHD computations.  Furthermore, unlike a
standard disk with constant-$\alpha$ viscosity, the disk midplane does
not cool off over time, though the surface cools as the star evolves
along the Hayashi track. Instead, the MRI may pile material in the dead
zone, causing it to heat up over time.  The ice line is firmly in the
terrestrial planet-forming region throughout disk evolution and can move
either inward or outward with time, depending on whether pileups form
near the star. Finally, steady-state mass transport is
an extremely poor description of flow through an MRI-active disk, as we
see both the turnaround in the accretion flow required by conservation
of angular momentum and peaks in $\dot{M}(R)$ bracketing each side of
the dead zone. We caution that MRI activity is sensitive to many
parameters, including stellar X-ray flux, grain size, gas/small grain
mass ratio and magnetic field strength, and we have not performed an
exhaustive parameter study here. Our 1+1D model also does not
include azimuthal information, which prevents us from modeling
the effects of Rossby waves.
\end{abstract}

 \section{Introduction} \label{sec:intro}

A typical, low-mass T-Tauri star accretes mass at a rate of
~$10^{-8}\;\mathrm{M_{\sun}\;yr^{-1}}$---one Jupiter mass every 100,000
years \citep[e.g.][]{Hartmann98, Sicilia04}. \citet{BH91} put forth the
magnetorotational instability (MRI) as the most likely driver of
accretion in T-Tauri disks, followed by \citet{brandenburg95},
\citet{Hawley95}, \citet{BHS96}, and \citet{bh98}. Before a
first-principles physical description of angular momentum transport was
available, accretion was often modeled using the $\alpha$-prescription \citep{SS73},
\begin{equation}
\nu = \alpha c_s H.
\label{eq:alpha}
\end{equation}
The $\alpha$-prescription relates turbulent viscosity to length
and velocity scales in the disk based on dimensional analysis.
In Equation \ref{eq:alpha}, $\nu$ is the turbulent
viscosity, $c_s$ is the sound speed and $H$ is the pressure scale
height. $\alpha$ is a
dimensionless efficiency that is often 
assumed, without physical motivation, to be uniform throughout the disk.






The first numerical investigations of MRI-driven turbulence were local
shearing box simulations \citep{Hawley95}, which treat a box of
approximate size $2 \pi H \times H \times H$ centered at a given
distance from the star. The shearing box is still a useful technique for
investigating detailed properties of turbulence. 
However, more recent investigations computed turbulent 
viscosity from first principles for global disk models, which were 
previously the domain of the $\alpha$-prescription. Such global 
simulations confirmed that the MRI
can lead to turbulence-driven accretion matching observed rates of
$10^{-8}\;{\mathrm{M_{\sun}\;yr^{-1}}} $ in either unstratified disks
\citep{hawley01, steinacker02, lyra08} or thin, stratified disks
\citep{sorathia10}. MRI turbulence is self-sustaining for the simulation
timeframe of around 1000 orbits near the inner boundary \citep{FN06,
flock11}.

In parallel with the development of global accretion disk models came
investigations of the behavior of the MRI in non-ideal, partially
ionized fluids. \citet{Gammie96} was the first to point out that
protostellar disks have surface layers ionized well enough to
couple to magnetic fields, and an interior dead zone where
extremely low ionization levels prevent magnetically driven
turbulence. Subsequent investigations
incorporating Ohmic resistivity into the MHD
equations confirmed Gammie's prediction that a high enough resistivity
would dampen the growth of the MRI \citep{Jin96, sano99}.  Numerical
simulations showed a ``dead zone'' near the midplane and near the star,
where the surface density is high enough to shield the disk interior
from ionizing radiation \citep{Fleming2000, sano00}. Recent research has
revealed that the dead zone is not entirely dead: the shielded interior
still experiences a shear stress only about an order of magnitude less
than the active layer due to propagating acoustic waves \citep{FS03} and
smooth, large-scale magnetic fields \citep{Turner07, TS08}.  Other
non-ideal effects that affect the MRI include the Hall current
\citep[e.g.][]{wardle99, balbus01, SS02} and ambipolar diffusion
\citep[e.g.][]{blaes94, maclow95, kunz04, BS11} (See Section
\ref{subsec:nonidealMRI} for a more complete description of each
effect).  Improved understanding of both non-ideal MRI and ionization in
disks \citep{igea99, IN06} has provided the tools to describe
the MRI in gas of any ionization fraction or density that can be found
in a protostellar disk.

Detailed short-timescale snapshots have now been constructed of angular
momentum transport in a protostellar disk, including the dead zone,
turbulent layers and corona \citep{dzyurkevich10, KL10, Bai11, flaig12}.
Yet it is possible that MRI-active T-Tauri disks may not be in steady
state due to (a) the dead zone and its resulting, radially varying
accretion rate and (b) the lack of a protostellar envelope to provide
material to maintain steady, inward mass transport. Our protostellar
disk snapshot, then, must change over the million-year timescales on
which the star/disk system evolves. The goal of this work is to
illustrate how disk interior structure and angular momentum transport
change over the entire, multi-million-year lifetime of the T-Tauri
phase. We extend the work of \citet{armitage01} and \citet{zhu10}, who
also performed million-year simulations, by calculating a radially and
vertically varying $\alpha$ based on the ionization state rather than
assuming a constant $\alpha$-value in the active zone.
\citet{martin12b} also modeled FU Orionis outbursts using time-dependant
global simulations of MRI-active disks, including Ohmic resistivity,
using 1-D layered models which used the
$\alpha$-prescription for the active layer, and an analytical
approximation for the active layer surface density \citep{martin12a}.
Our models build on their work by adding a vertical dimension to
the disk structure.

Our evolving model of a magnetically turbulent T-Tauri disk answers the following questions:

\begin{enumerate}

\item How do the relative sizes of the dead zone and active
layers change over time?

\item How does $\dot{M}$ vary with radius and time?

\item How can disk modelers parameterize heating in the active
layers and dead zone without resorting to a 3-D MHD simulation?

\item Does the disk midplane heat up or cool off over time?

\item Where is the ice line in an MRI-active disk and how does
its location change over time?

\end{enumerate}
Questions 1, 2 and 3 elucidate basic properties of an MRI-turbulent
accretion disk. Questions 4 and 5 highlight fundamental ways in which
disk models based on the standard viscosity prescription with constant
$\alpha$ lead us astray.  Note that our disk model does not include
photoevaporation, which is another process that operates on million-year
timescales that can produce radially varying accretion accretion rates.


Our ability to simulate an entire T-Tauri disk lifetime is due to a new
MRI activity prescription that allows us to compute the thermal and
viscous effects of MRI turbulence without resorting to 3-D
magnetohydrodynamic simulations of the turbulence itself. We can thus
reduce our computational domain from three spatial dimensions to 1+1
spatial dimensions---1-D vertical structures representing axisymmetric
disk annuli that are connected only by a 1-D radial mass transport
equation \citep{DR09}. Sacrificing information about small-scale
turbulent fluctuations, we retain our ability to accurately describe
large-scale structures such as the dead zone and active layers while
dramatically improving our ability to simulate long timescales.


Our paper is organized as follows. In Section \ref{sec:MRI} we discuss
the basic equations governing the MRI under non-ideal MHD conditions and
give our prescription for determining turbulent viscosity. We present
our method of computing vertical strtucture and mass transport in
Section \ref{sec:model}. We outline a basic picture of MRI-turbulent
disk evolution and answer Questions 1, 2 and 3 in Section
\ref{sec:mass_transport}. In Section \ref{sec:thermo}, we discuss the
differences in thermal structure between our model and constant-$\alpha$ disk models, which leads us to answer Questions 4 and 5. We
discuss the limitations of our model in Section
\ref{section:limits} and present our conclusions in Section \ref{sec:conclusions}.  Readers who
wish to skip over the details of the computations may wish to
proceed directly to Section \ref{sec:mass_transport}.


\section{Simulating Magnetorotational Instability-Driven Turbulence in
Partially Ionized Gases}
\label{sec:MRI}

When an accretion disk is fully ionized  and the magnetic field is weak,
the entire disk is MRI turbulent.  Yet when a disk is only partially
ionized, as is the case for a protoplanetary disk, there is an
incomplete coupling between the disk gas and the magnetic field, and
non-ideal effects become important to the growth of MRI-driven
turbulence. In this section, we describe how we treat angular momentum
transport from MRI turbulence in non-ideal, partially ionized gases. We
begin by describing our turbulence criterion in Section
\ref{subsec:nonidealMRI}, then list our method for determining the
diffusion regime and resulting turbulent stress in Section
\ref{subsec:regime}.

\subsection{MRI activity criteria in the three non-ideal regimes}
\label{subsec:nonidealMRI}

The non-ideal magnetic induction equation has three extra terms
corresponding to the three non-ideal effects \citep{Wardle07}, Ohmic
resistivity, the Hall effect and ambipolar diffusion:
\begin{equation} \label{eq:mag_diff}
\begin{split}
\frac{\partial \boldsymbol{B}}{\partial t} =
  \nabla\times(\boldsymbol{v}\times\boldsymbol{B}) -  \\
  \nabla\times\left[\eta_{O}\nabla\times \boldsymbol{B} + 
  \eta_{H}(\nabla\times \boldsymbol{B})\times\boldsymbol{\hat{B}} +
  \eta_{A}(\nabla\times \boldsymbol{B})_{\bot}\right].
\end{split}
\end{equation}
In Equation \ref{eq:mag_diff}, $\boldsymbol{v}$ is the gas velocity,
$\boldsymbol{B}$ and $\boldsymbol{\hat{B}}$ are the magnetic field and
magnetic field unit vector, and $\bot$ refers to the component
perpendicular to $\boldsymbol{B}$. $\eta_{O}$, $\eta_{H}$, and
$\eta_{A}$ are the Ohmic, Hall, and ambipolar diffusivities
respectively.  Ohmic resistivity dominates other non-ideal effects when
collisions with neutrals cause both electrons and ions to decouple from
field lines. When collisional drag is sufficient to decouple ions and
grains from the magnetic field, but not electrons, the relative velocity
between the ions and electrons is non-negligible and the Hall effect is
dominant. In the ambipolar diffusion regime, electrons and ions decouple
from the neutral gas and the magnetic field lines are frozen to the
charged species and drift through the neutral gas.

When Ohmic resistivity is the largest non-ideal effect, the MRI will
only occur if the Elsasser number,
\begin{equation} \label{eq:Elsasser}
\Lambda\equiv \frac{v_{Az}^{2}}{\eta_{O}\Omega},
\end{equation}
is at least of order unity \citep{SS02,Turner07}. In Equation
(\ref{eq:Elsasser}), $v_{Az}$ is the Alfv$\mathrm{\acute{e}}$n speed in
the vertical direction and $\Omega$ is the Keplerian angular velocity.
Physically, $\Lambda$ is the ratio of the wavelength of maximum
growth to the diffusive scale length.
The tangled magnetic fields in MRI turbulence usually
have a toroidal component with pressure 10 to 30 times greater than the
pressure in the vertical component \citep{TS08}, so the Alfv\'{e}n speed
in the vertical direction used to calculate $\Lambda$ is
\begin{equation}
v^2_{Az} \sim \frac{1}{10} v^2_A,
\label{eq:vertalfven}
\end{equation}
where $v_A = B / \sqrt{4 \pi \rho}$ is the total Alfv\'{e}n
speed.
The $\Lambda \ga 1$ criterion ensures that the most unstable mode can
grow more quickly than the charged particles can diffuse across magnetic
field lines. Simulations by \citet{SS02} suggest the Hall effect does
not change the conditions for turbulence if Ohmic diffusion is also
present. Further work is required to determine the growth of the MRI in
regimes where the Hall term is much stronger than other terms
\citep{wardle12}.


Ambipolar diffusion arises from the relative motion of ions and neutral
particles in the disk gas. In the ``strong coupling'' limit, in which
ion density is negligible and electron recombination time is much
smaller than the orbital time $1/\Omega$, as is generally the case
for protoplanetary disks \citep{Bai11}, ion density cannot be assumed to
follow the continuity equation. Instead, the ion density is determined
by the ionization-recombination equilibrium, and characterized by the
parameter $Am$ \citep{CM07}:
\begin{equation}
Am\equiv \frac{\gamma\rho_{i}}{\Omega},
\label{eq:Am1}
\end{equation}
where $\rho_{i}$ is the ion density and $\gamma$ is the neutral-ion
drag coefficient,
\begin{equation}
\gamma = \frac{\left\langle \sigma_{ni}
w_{ni}\right\rangle}{m_{n}+m_{i}}.
\label{eq:dragcoeff}
\end{equation}
In Equation \ref{eq:dragcoeff}, $\sigma_{ni}$ is the effective cross
section for neutral-ion collision and $w_{ni}$ is the relative velocity
between neturals and ions.  Physically, $Am$ is the ratio of the orbital
period to the collisional timescale between ions and neutrals. Since
$\eta_A=v_{A}^{2}/\gamma\rho_{i}$ \citep{BS11}, one can rewrite Equation
\ref{eq:Am1} as
\begin{equation} \label{eq:Am}
  Am=\frac{v_{A}^{2}}{\eta_{A}\Omega},
\end{equation}
which is equivalent to the Elsasser number $\Lambda$ in the Ohmic
regime.  Similarly then, $Am$ is the ratio of the wavelength of
maximum growth to
the ambipolar diffusive scale length.

In their three-dimensional shearing-box simulations exploring the effect
of ambipolar diffusion on MRI turbulence, \cite{BS11} determine that
heavily ionized yet tenuous disks can only sustain turbulence when
threaded by weak magnetic fields. The magnetic field strength is
characterized by the plasma $\beta$, the ratio of the gas pressure to
the magnetic pressure:
\begin{equation}
\beta = \frac{8\pi P}{\left|\boldsymbol{B}\right|^{2}}.
\label{eq:beta}
\end{equation}
The requirement that the magnetic field energy be small in comparison to
the gas thermal energy (a ``weak'' field) restricts MRI turbulence to
values of the plasma $\beta$ that are greater than a minimum \citep{BS11}:
\begin{equation}
\label{eq:betamin}
  \beta_{min}(Am) = \left[ \left(\frac{50}{Am^{1.2}}\right)^{2} +
  \left(\frac{8}{Am^{0.3}}+1\right)^{2}\right]^{1/2}.
\end{equation}
The maximum field strength, beyond which the field is too strong to be
destabilized for any given field geometry, decreases the more important
ambipolar diffusion becomes (smaller $Am$). However, for a sufficiently
weak field, MRI can be sustained even for $Am\ll 1$.

In a protoplanetary disk, ambipolar diffusion dominates in the
atmosphere, which is diffuse and highly ionized by stellar X-rays and
cosmic rays.  Ambipolar diffusion is also important in the outer disk
where the surface density is very low. Ohmic resistivity dominates in
the dense inner region shielded from ionizing radiation. Ignoring the
effects of the Hall diffusivity, which are unlikely to alter either the
conditions required for MRI or the strength of the turbulence where
Ohmic dissipation is also present \citep{SS02}, a non-ideal
protoplanetary disk differs from an ideal accretion disk through the
possibility of a dead zone in the inner disk. The dead zone would
remain cold and would not efficiently transport angular momentum.  The
accretion efficiency in the upper, ionized layers of non-ideal disks
also lags behind ideal disks due to ambipolar diffusion effects.

In order to compute the disk viscosity and angular momentum transport
properties, we need to know (a) whether the MRI is operating, and (b),
if so, how strong the turbulence is. Since the MRI growth
timescale is roughly $1/\Omega$, hundreds of thousands of times
shorter than our $\sim 1$~Myr simulation timescale, we assume
the MRI is either fully saturated or completely damped. Our
MRI turbulence criterion is therefore equivalent to that of
\citet{Bai11}:
\begin{itemize}
\item If $\Lambda \geq 1$ and $\beta > \beta_{min}$, 
neutral gas couples to the magnetic field and MRI is saturated;
\item If $\Lambda < 1$ or $\beta \leq \beta_{min}$, neutral gas
decouples from the magnetic field and MRI is damped.
\end{itemize} 
In the next section, we discuss the computation of the magnetic
diffusivities that determine $\Lambda$ and $Am$.


\subsection{The diffusion regime and turbulent stress}
\label{subsec:regime}

We cannot apply our turbulence criterion without knowing the values of
$\eta_O$, $\eta_H$ and $\eta_A$. For a given charged species $j$, the
ratio of Lorentz force to the neutral drag force is
\begin{equation}
\beta_j = \frac{Z_j e B}{m_j c \gamma_j \rho},
\label{eq:lorentz_drag}
\end{equation}
where $Z_j e$ is the charge of $j$ (negative or positive), $B$ is the
magnitude of the magnetic field, $m_j$ is the mass of $j$, $c$ is the
speed of light, and $\gamma_j$ is defined according to Equation
\ref{eq:dragcoeff}. (Note that $\beta_j$ is not the same as the plasma
$\beta$ of Equation \ref{eq:beta}.) For each diffusion regime, one can
define a conductivity by summing over all charged species:
\begin{align}
& \sigma_O = \frac{ec}{B} \sum_j n_j Z_j \beta_j, \\
& \sigma_H = \frac{ec}{B} \sum_j \frac{n_j Z_j}{1 + \beta_j^2}, \\
& \sigma_P = \frac{ec}{B} \sum_j \frac{n_j Z_j \beta_j}{1 +
\beta_j^2}
\label{eq:conductivities}
\end{align}
\citep{Wardle07}. In Equations 11-\ref{eq:conductivities}, $n_j$ is the
number density of species $j$. Finally, one can write the diffusivities
according to
\begin{align}
& \eta_O = \frac{c^2}{4 \pi \sigma_O}, \\
& \eta_H = \frac{c^2}{4 \pi \sigma_{\bot}}
\frac{\sigma_H}{\sigma_{\bot}}, \\
& \eta_A = \frac{c^2}{4 \pi \sigma_{\bot}}
\frac{\sigma_P}{\sigma_{\bot}} - \eta_O,
\label{eq:diffusivities}
\end{align}
where $\sigma_{\bot} \equiv \sqrt{\sigma_H^2 + \sigma_P^2}$.

To determine the equilibrium abundances of charged species
$n_j$, we solve a simplified set of chemical reactions from
Model 4 of \citet{IN06}, which we briefly motivate here.
The set is derived from the following reactions:
\begin{align}
{\mathrm H}_2 + {\rm X} &\rightarrow {\mathrm H}_2^+ + e^- \\
{\mathrm H}_2^+ + {\mathrm H}_2 &\rightarrow {\mathrm H}_3^+ +
{\mathrm H} \\
{\mathrm H}_3^+ + {\rm CO} &\rightarrow {\rm HCO}^+ + {\mathrm H}_2
\\
2 {\mathrm H} + {\mathrm g} &\rightarrow {\mathrm H}_2 + {\mathrm
g} \\
{\rm HCO}^+ + e^- &\rightarrow {\rm CO} + {\mathrm H} \\
{\rm HCO}^+ + {\rm Mg} &\rightarrow {\rm Mg}^+ + {\rm CO} +
{\mathrm H} \\
{\rm Mg}^+ + e^- &\rightarrow {\rm Mg},
\label{eq:reactions}
\end{align}
where HCO$^+$ is a representative molecular ion, Mg$^+$ is a
representative metal ion and g is a grain. Here every species (except the
energetic particle X---a cosmic ray, X-ray or radionuclide decay
product) is created in at least one reaction, and
destroyed in at least one other. Over the whole set, no species
is produced or consumed on balance.
The subset producing the ions and electrons reduces to 
\begin{equation}
2 {\mathrm H}_2 + 2 {\rm X} + 2 {\rm CO} \rightarrow {\rm H}_2 +
2 {\rm HCO}^+ + 2 e^-.
\label{eq:reducedion}
\end{equation}
That is, each energetic particle striking a hydrogen molecule
yields one ion and one electron. In constructing the
conductivity lookup tables we therefore approximate Equations
17-21 by
\begin{align}
{\rm H}_2 + {\rm X} &\rightarrow {\rm HCO}^+ + e^- \\
{\rm HCO}^+ + e^- &\rightarrow {\rm H}_2,
\label{eq:approxion}
\end{align}
neglecting the fact that the molecular ion contains just one
hydrogen atom. Since HCO$^+$ is orders of magnitude less
abundant than H$_2$, forming ions leaves the H$_2$ density
unchanged. Similarly, we don't model CO destruction and
reformation because the ion is so much less abundant than the
molecule. Equation 22 then becomes
\begin{equation}
{\rm HCO}^+ + {\rm Mg} \rightarrow {\rm Mg}^+ + {\rm H}_2.
\label{eq:hcoplus}
\end{equation}
The simplified network consists of Eqs. 25, 26, 27 and 23, together
with the grain surface reactions described by \citet{IN06}. The
metal atoms' thermal adsorption and desorption on the grains is
included. \citet{IN06} found that this reduced network yields
similar results to a detailed version including hundreds of
species and thousands of reactions, in the most common situation
where the recombination occurs mostly on the grains.

The internal grain density, gas/small grain ratio, and grain size used
in the chemical reaction network are listed in Table
\ref{table:diskpars}. Here we deviate from the standard interstellar
gas/small grain mass ratio of 100 and assume some grain growth has
occurred, so that 90\% of the grain mass is in grains larger than $1
\mu$m. Using the standard gas/dust ratio of 100 resulted in no MRI
turbulence (see section \ref{sec:turbulent_structure}). To avoid
having to run the chemical reaction network at every timestep of our
million-year simulations, we followed the approach of \citet{flaig12}
and created a look-up table of magnetic diffusivities as a function of
temperature $T$, gas density $\rho$, ionization rate $\zeta$ and plasma
$\beta$. To generate the look-up table, we ran the chemical reaction
network until it reached equilibrium abundances of all species for each
combination of $T$, $\rho$, $\zeta$ and $\beta$. We then computed the
conductivities and tabulated diffusivities according to Equations 10-16.

\begin{figure*}[ht]
\centering
\begin{tabular}{cc}
\includegraphics[trim=1.5cm 0cm 0cm 0cm, clip=true, width=0.48\textwidth]{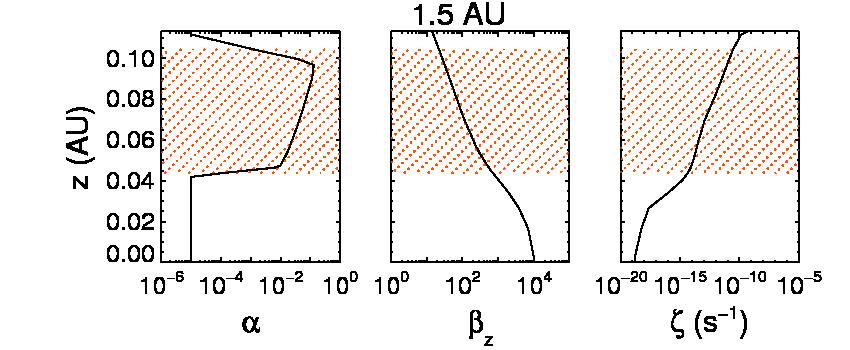}
\includegraphics[trim=1.5cm 0cm 0cm 0cm, clip=true, width=0.48\textwidth]{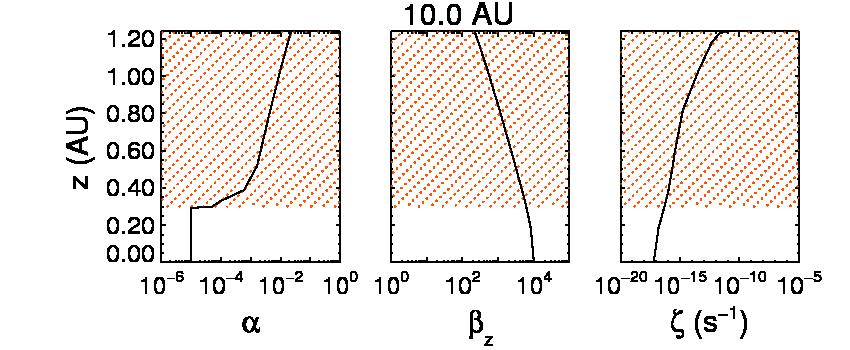} \\
\includegraphics[trim=1.5cm 0cm 0cm 0cm, clip=true, width=0.48\textwidth]{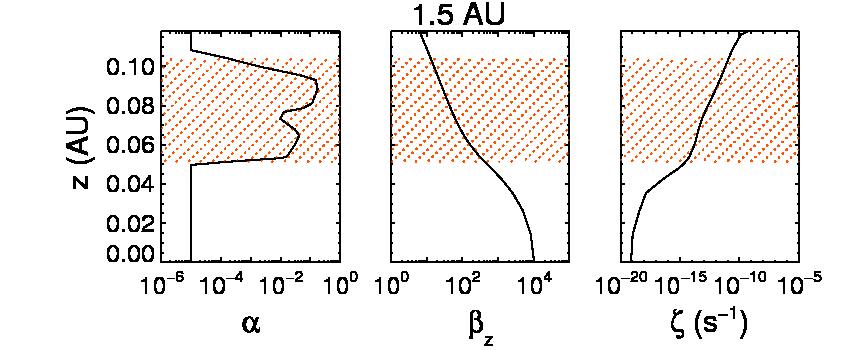}
\includegraphics[trim=1.5cm 0cm 0cm 0cm, clip=true, width=0.48\textwidth]{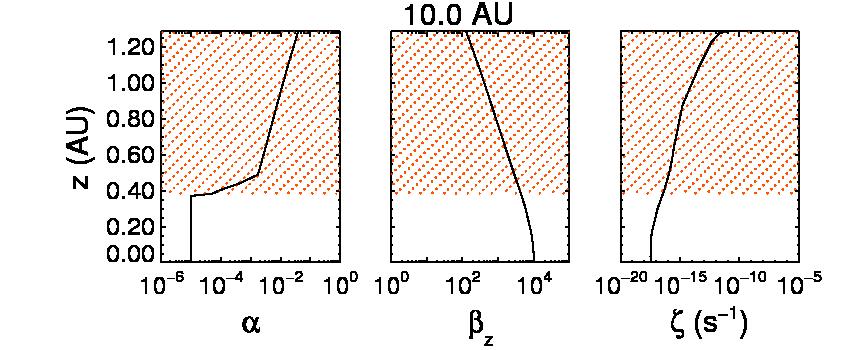}
\end{tabular}
\caption{Plots showing the multi-layered structure of evolved,
magnetically active disks. Besides the dead zone and the active layer,
both disks have a corona near the star and at the surface where the
magnetic field is too strong for the tenuous gas to create turbulence.
Disks may also have a double-layered active zone with a dead
slice in some locations. {\bf
Top left}: $\alpha(z)$, $\beta_z(z)$ and $\zeta(z)$ of Model 1 at 1.5~AU
after 1~Myr. {\bf Top right}: Model 1 at 10~AU. {\bf Bottom left}: Model
2 at 1.5~AU, 1 Myr. {\bf Bottom right}: Model 2 at 10~AU. In each panel,
the active region with $\Lambda > 1$ and $\beta > \beta_{min}$ is shaded
with red dots.}
\label{fig:vstructline}
\end{figure*}

The rate coefficient for the reaction $\mathrm{H}_2 + {\rm X} \rightarrow
\mathrm{H}_2^+ + e^-$ is, of course, the ionization rate $\zeta$.  To
determine $\zeta$, we consider cosmic rays, stellar X-rays, and
short-lived radionuclides ($\tau_{1/2} \ll 10^{8}$ yr), including
$\mathrm{^{26}Al}$.  Following \citet{UN09}, we take the ionization rate
from short-lived radionuclides as $\zeta_{R}=7.6 \times 10^{-19}
\;\mathrm{s^{-1}}$ and calculate the attentuated cosmic ray ionization
as:
\begin{equation}
  \begin{split}
     \zeta_{CR}(z) = \frac{\zeta_{CR}^{surf}}{2}
     \left\{exp\left(-\frac{\Sigma_{1}(z)}{\lambda_{CR}}\right)
     \left[1+\left(\frac{\Sigma_{1}(z)}{\lambda_{CR}}\right) ^ {-3/4}
       \right]^{-4/3}\right.
     \\
     \left. +exp\left(-\frac{\Sigma_{2}(z)}{\lambda_{CR}}\right)
     \left[1+\left(\frac{\Sigma_{2}(z)}{\lambda_{CR}}\right) ^ {-3/4}
       \right]^{4/3}\right\},
  \end{split}
\label{eq:cr}
\end{equation}
where $\zeta_{CR}^{surf}$ is the unattenuated cosmic ray ionization
rate, $\lambda_{CR}$ is the cosmic ray penetration depth \citep{UN81},
and $\Sigma_{1, 2}(z)$ are the mass columns above and below the vertical
height $z$. Values of $\zeta_{CR}^{surf}$, $\lambda_{CR}$ and all other
numerical inputs to our model are listed in Table \ref{table:diskpars}.
Finally, following \citet{BG09}, we calculate the stellar X-ray
ionization rate $\zeta_{X}$ as:
   \begin{equation}
    \begin{split}
     \zeta_{X}(z)=L_{X,29}\left(\frac{R}{1
     \mathrm{AU}}\right)^{-2.2}\left\{ \zeta_{1}\left[
     e^{-\left(\frac{\Sigma_{1}(z)}{\lambda_{1}}\right)^{p_{1}}} +
     e^{-\left(\frac{\Sigma_{2}(z)}{\lambda_{1}}\right)^{p_{1}}}\right]
     + \right. \\ \left.
     \zeta_{2}\left[e^{-\left(\frac{\Sigma_{1}(z)}{\lambda_{2}}\right)^{p_{2}}}
     +
     e^{-\left(\frac{\Sigma_{2}(z)}{\lambda_{2}}\right)^{p_{2}}}\right]
     \right\},
    \end{split} \label{eq:xray}
   \end{equation}
where $L_{X,29}\equiv L_{X}/(10^{29}\;\mathrm{erg\;s^{-1}})$, and
$L_{X}$ is the the stellar X-ray luminosity. We take $L_{X,29}=20$ to
match the young solar-mass stars observed in the Orion Nebula
\citep{garmire00}.  Here we keep the stellar X-ray flux constant in
time, though it could certainly vary in either a smooth, systematic way
with age or stochastically with accretion bursts.  All parameters in
Equation \ref{eq:xray} are listed in Table \ref{table:diskpars}. The
first exponential represents attenuation of X-rays by absorption, and
the second represents the contribution from scattered X-rays. We show
vertical profiles of the ionization rate at two different disk radii in
Figure \ref{fig:vstructline}.

After the ionization rate is determined at each $(R,z)$ zone in our disk
model, an interpolation through the look-up table can return the
magnetic diffusivities. We can then compute the Elsasser number $\Lambda$
and $Am$ and determine whether MRI turbulence in the zone is active or
not, according to our turbulence criterion (Section
\ref{subsec:nonidealMRI}). The last remaining ingredient in our MRI
prescription is a rule for determining the strength of the turbulence,
where it is present. Where MRI is saturated, we use the scaling relations
\begin{align}
\alpha &= \frac{1}{2 \beta} \\
\nu &= \alpha c_s H = \frac{B^2}{16 \pi \rho \Omega}
\label{eq:alphascale}
\end{align}
found between turbulent stress and magnetic field strength in a
variety of shearing-box simulations \citep{Hawley95, Sano04, BS11}. Note
that Equation \ref{eq:alphascale} applies no matter the field
geometry or value of $\Lambda$ or $Am$. 

Since there is an accretion flow caused by large-scale magnetic fields
even in the dead zone \citep{TS08, nelson10}, we set a minimum value of
$\alpha$ where the MRI is damped. In the active layer, $\alpha$ is close
to its maximum value of $\sim 0.5$, set by the cessation of MRI in the
strong-field limit of $\beta \geq 1$ \citep{Hawley95,bh98}.
\citet{BS11} found a similar result in the ambipolar regime,
$\alpha\approx0.4$ for $Am\rightarrow\infty$. The shear stress in the
dead zone is an order of magnitude less than in the active layer
\citep{FS03,Turner07,TS08}, and the plasma $\beta$ at the midplane is
typically two to three orders of magnitude higher than at the top of our
grid (see Figure \ref{fig:vstructline}, which shows profiles of the
vertical component of $\beta$, $\beta_z = 8 \pi P / B_z^2$, for two
different disk radii). $\alpha_{min}$ must be therefore be roughly four
orders of magnitude less than $\alpha_{max}$ to produce an appropriate
level of dead zone shear stress. We take $\alpha_{min}=10^{-5}$ (Table
\ref{table:diskpars}).


\section{Disk Structure and Mass Transport}
\label{sec:model}

With our method of computing turbulent viscosity at any location in a
protostellar disk, we may now simulate how the disk re-distributes its
mass throughout its multi-million-year existence. Since our viscosity
prescription depends on the four inputs $T$, $\rho$, $\beta$ and
$\zeta$, which are functions of radius and vertical height $(R,z)$, we
must compute the detailed vertical and radial structure of the disk. Our
computational setup, similar to the disk models of \citet{DR09}, is
based on the following simplifying assumptions:
\begin{enumerate}
\item The disk is axisymmetric and symmetric about the midplane;
\item The disk is geometrically thin, $H / R \ll 1$;
\item Heat escapes in the vertical direction much faster than
it is carried with the gas flow in the radial direction.
\end{enumerate}

Assumption 1 reduces the physical three-dimensional disk to a
two-dimensional quadrant with zero flux at the midplane (required for
symmetry). Neglecting the azimuthal dimension is a critical step in
speeding up the code to allow long-timescale simulations. Assumption 2
allows the vertical and radial dimensions to be decoupled in a 1+1-D
framework, so that energy transport proceeds only in the vertical
direction. Each radial gridpoint contains an independent vertical
structure model. Assumption 2 is valid as long as the vertical
sound-crossing time (of order the orbital timescale) is much less than
the accretion timescale, which is generally true in T-Tauri disks.

In section \ref{subsec:vstruct} we describe our vertical structure
model, while in Section \ref{subsec:diffusion} we discuss our mass
transport parametrization. Section \ref{subsec:methods} contains a
description of our computational methods.


\subsection{Vertical structure}
\label{subsec:vstruct}

Hydrostatic equilibrium and the thermal balance between stellar
irradiation, radiative cooling and viscous heating govern the vertical
structure of our disk. We use the flux-limited diffusion approximation
for the transport of viscously generated energy. The accretion flux
gradient is determined by the viscous energy generation rate per unit
volume \citep{Pringle81}:
\begin{equation} \label{eq:visc_energy}
\frac{\partial F_{acc}}{\partial z}=\frac{9}{4}\nu\Omega^{2}\rho.
\end{equation}
In Equation \ref{eq:visc_energy}, $\Omega$ is the height-dependent
Keplerian frequency
\begin{equation}
\Omega = \left [ \frac{GM_*}{\left ( R^2 + z^2 \right )^{3/2}}
\right ]^{1/2} .
\label{eq:Omega}
\end{equation}
Temperature and pressure are related by the ideal gas equation of state,
\begin{equation} \label{eq:ideal}
P=\left(\frac{R_{g}}{\mu}\right)\rho T,
\end{equation}
where $\mu = 2.33$~g~mol$^{-1}$ is the mean molar weight of the disk gas
(Table \ref{table:diskpars}).  The temperature and pressure gradients
are
\begin{equation} \label{eq:rad_trans}
\frac{\partial T_{acc}}{\partial z} = \frac{\partial P}{\partial
z}\nabla\frac{T}{P},
\end{equation}
\begin{equation} \label{eq:hydro_eq}
\frac{\partial P}{\partial z} = -\rho\Omega^{2}z,
\end{equation}
In Equation \ref{eq:rad_trans}, $T_{acc}$ is the temperature contribution
from viscous heating only---stellar irradiation and the ambient
molecular cloud also contribute some of the thermal energy. $T$
is the true temperature and includes all heat sources.

To calculate $\nabla \equiv d \ln T / d \ln P$, the thermodynamic
gradient, we use the Schwarzschild criterion for stability against
convection:
\begin{equation}
\nabla = \left \{ \begin{array}{l} \nabla_{rad}, \nabla_{rad}
\leq \nabla_{ad} \\
\nabla_{conv}, \nabla_{rad} > \nabla_{ad} \end{array} \right. .
\label{eq:nabla}
\end{equation}
$\nabla_{ad} = 2/7$ is the adiabatic thermodynamic gradient for diatomic
gas and
\begin{equation} \label{eq:nabla_rad}
\nabla_{rad}=\frac{3}{4}\frac{\kappa PF}{ac\Omega^{2}zT^{4}},
\end{equation}
is the radiative thermodynamic gradient, where $a$ is the radiation
density constant and $\kappa$ is the local Rosseland mean opacity. For
full details on how to compute $\nabla_{conv}$ in the disk's convective
zone, see \citet{KW94}. Computing $\nabla$ requires the local Rosseland
mean opacity, $\kappa$. At low temperatures, $T < 700$~K, we use the
opacities of \citet{Semenov03} calculated for a 5-layered sphere
topology. At higher temperatures where molecular gas dominates opacity
($T > 1000$~K), we use the tables of \citet{Ferguson05}.  For $700 \; {\rm
K} \leq T \leq 1000$~K, we interpolate between the two tables using a
weighted average in $\log (T)$ space.  For more details, and plots of the
resulting opacities, see \citet{DR09}.

Integrating the coupled ODEs in Equations 32, 35 and 36 requires
two different temperatures: $T$, the true temperature resulting from all
sources of thermal energy, and $T_{acc}$, the component from viscous
heating only.  The other heat sources are the central star, which sets
an equilibrium temperature component $T_{eq}$, and the ambient
star-forming region, from which long-wavelength radiation that
penetrates the disk sets a minimum temperature $T_{amb}$. To compute
$T_{eq}$, we begin by assuming the disk surface is flared as a result of
hydrostatic equilibrium and radiative equilibrium with the star.
Following the models developed by \citet{CG97}, we calculate the grazing
angle $\theta$ at which stellar energy enters the disk:
\begin{equation}
\theta\approx\frac{8}{7}\left(\frac{T_{*}}{T_{c}}\right)^{4/7}
\left(\frac{R}{R_{*}}\right)^{2/7},
\label{eq:grazing}
\end{equation}
where $R_*$ and $T_*$ are the star's radius and effective temperature
and $T_c$ is a measure of the gravitational potential at the surface of
the star:
\begin{equation}
T_c=\frac{GM_{*}\mu}{\sigma R_{*}}.
\label{eq:Tc}
\end{equation}
In Equation \ref{eq:Tc}, $M_{*}$ is the star's mass and $\sigma$ is the
Stefan-Boltzmann constant.  As the disk evolves, we determine the star's
temperature and radius as a function of age from the pre main-sequence
evolutionary tracks of \citet{DM94}. Since the star's luminosity
decreases as it moves down the Hayashi track, the disk flaring becomes
less pronounced over time and the disk surface, where $T \approx
T_{eq}$, cools.

In the direction parallel to the disk midplane, the stellar radiation
penetrates to an optical depth $\tau_*^{\|}\sim 1$. The asterisk subscript 
denotes that this optical depth is measured at the peak wavelength of
the starlight, near $1 \; \mu$m. Measured perpendicular to the disk
midplane, stellar radiation is mostly attenuated by optical depth
$\tau_*^{\bot}\sim\tau_*^{\|} \theta\sim\theta$. The equilibrium
temperature between the stellar heating and radiative cooling at the
disk surface---neglecting viscous heating---is \citep{Dalessio06,Natta00}:
\begin{equation}
T_{eq,surf}\approx 0.8\left( \frac{\theta}{\tau_{surf}}
\right)^{1/4} \left(\frac{R_{*}}{R}\right)^{1/2} T_{*},
\label{eq:teqsurf}
\end{equation}
where $\tau_{surf}$ is the Rosseland mean optical depth at the disk
surface for blackbody radiation at $T_{eq,surf}$. The top surface of our
vertical grid is defined by $\tau_{surf} = 0.2$. Since half of the
stellar radiation absorbed by grains at the disk surface is re-radiated
into space, the equilibrium temperature with the star as a function of
height in the disk---again, neglecting viscous heating---is
\begin{equation}
T_{eq}(z)^4 = \frac{1}{2} T_{eq,surf}^4 e^{-\tau_z},
\label{eq:teq}
\end{equation}
where $\tau_z$ is the Rosseland mean optical depth to height z:
\begin{equation}
\tau_z = \int_z^{z_{surf}} \kappa \rho \: dz.
\label{eq:tauz}
\end{equation}
Finally, the true temperature $T$ simply the flux sum of the individual
temperature components:
\begin{equation}
T^4 = T_{acc}^4 + T_{eq}^4 + T_{amb}^4.
\label{eq:ttrue}
\end{equation}

\subsection{Radial diffusion}
\label{subsec:diffusion}

Since the radial and vertical dimensions of our disk model are
decoupled, we must treat mass transport as a one-dimensional problem.
Yet the MRI-active, partially ionized disk is vertically layered, with
the most active accretion occurring at the surface. The key to a
successful 1-D description of layered accretion is the fact that
vertical re-distribution of mass within an annulus occurs more quickly
than accretion in the active layers: $1/\Omega \ll R^2/\nu_{active}$. By
computing a mass-weighted, vertically averaged value of turbulent
viscosity in each annulus,
\begin{equation}
\bar{\nu} = \frac{2}{\Sigma} \int_{z=0}^{z_{surf}} \nu \rho
dz,
\label{eq:nuave}
\end{equation}
where $\Sigma$ is the surface density in the annulus, we can
describe mass transport using the radial diffusion equation:
\begin{equation}
\label{eq:diffusion}
 \frac{\partial \Sigma}{\partial t} = \frac{3}{R}
 \frac{\partial}{\partial R} \left[R^{1/2}\frac{\partial}{\partial
 R}\left(\Sigma \bar{\nu} R^{1/2}\right)\right].
\end{equation}

In each $(R,z)$ zone, we compute viscosity according to Equation
\ref{eq:alpha}. We use a height-dependent modified scale height
$H$,
\begin{equation}
\label{eq:scaleheight}
H = \frac{c_{s}/\Omega}{\sqrt{1 + \left(2z^{2} \Omega^{2}/ c_{s}^{2}
\right)}},
\end{equation}
softened into a non-singular form \citep{Milsom94}. In MRI-active zones,
$\alpha$ is given by Equation \ref{eq:alphascale}, while in inactive
zones we set $\alpha$ to our chosen value of $\alpha_{min}$ (Table
\ref{table:diskpars}). The sound speed used to compute
$\nu(R,z)$ is
\begin{equation}
c_{s}^{2} = \frac{R_{g}}{\mu}T.
\label{eq:sound_speed}
\end{equation}

\subsection{Computational methods and initial conditions}
\label{subsec:methods}

To initialize the disk evolution model, we compute the vertical
structure of a disk with the following features:

\begin{enumerate}

\item A surface density profile $\Sigma\propto R^{-3/2}$, predicted by
\citet{zhu10} for layered accretion disks in the T-Tauri phase.

\item A pre-main-sequence star with a mass of 0.95 $M_{\sun}$ and an
initial age of 0.1 Myr, which roughly coincides with the beginning of
the T-Tauri phase \citep{dunham12}. The star will continue to accrete a
small amount of mass from the disk during the $\sim 3$~Myr T-Tauri
phase.

\item An ambient temperature of 20~K (Table
\ref{table:diskpars}) to match the typical background
temperatures of infrared dark clouds \citep{peretto10}.

\item An outer radius of 70~AU, set by the 80~AU solar nebula size limit
of \citet{Kretke12}. (Note that the disk expands from its initial
radius.) Kretke et al.\ show that a solar nebula with $R_{out} > 80$~AU
would excite Kozai oscillations in some of the planetesimals scattered
by Jupiter and Saturn, stranding them in stable, high-inclination,
low-eccentricity orbits that surveys have not detected.

\item An inner radius of $R_{in} = 0.5$~AU. The requirement that the
disk stay below the dissociation temperature of H$_2$, so that the ideal
gas equation holds, dictates our choice of $R_{in}$. \citet{RL86} find
that the exact value of $R_{in}$ does not affect the overall disk
structure as long as $R_{out} \gg R_{in}$.

\item A disk mass of either $0.015 M_{\odot}$ (Model 1) or $0.03
M_{\odot}$ (Model 2). Our disk masses are designed to be comparable to
previous global-disk MRI simulations \citep[e.g.][]{lyra08,
dzyurkevich10, Bai11}, almost all of which use minimum-mass solar
nebulae. Note that such low-mass disks are probably not viable giant
planet-forming environments \citep[e.g.][]{thommes08}. In a forthcoming
study, we will examine the evolution of MRI-active, high-mass,
planet-forming disks.

\end{enumerate}

We then evolve the disk forward in time using Equation
\ref{eq:diffusion}. We use fully implicit finite differencing, adjusting
the timestep $\Delta t$ so that surface density varies by a maximum of
1.0\% during a single timstep. The inner boundary $R_{in}$ experiences
zero stress, such that matter falls directly from $R_{in}$ onto the
star. The disk is allowed to expand freely from the outer boundary
$R_{out}$, with four new zones added to the disk each diffusion time of
$R_{out}^2 / \bar{\nu}(R_{out})$.

At each timestep, we independently calculate the vertical structure for
each zone in the radial grid. We begin the vertical structure solver
with initial guesses of $T_{acc}$, $\rho$ and vertical component of the
magnetic pressure $P_{B,z}=B_{z}^2/8\pi$ at the top of our grid, defined
by $\tau_{surf} = 0.2$. We find the height of the grid surface by
\begin{equation}
z_{surf} = \frac{\kappa(\rho,T) P}{\Omega^2 \tau_{surf}}
\label{eq:zsurf}
\end{equation}
The accretion flux at the grid surface is $F_{acc}(z_{surf}) = \sigma
T_{acc}^4$.  We use a fourth-order Runge-Kutta integrator with adaptive
stepsize control to integrate the coupled ODEs in Equations
\ref{eq:visc_energy}, \ref{eq:rad_trans} and \ref{eq:hydro_eq} from the
surface to the midplane. The vertical magnetic pressure stays
constant in height, though it varies with radius. A solved
vertical structure model has the properties
\begin{equation}
F_{acc}(z=0) = 0,
\label{eq:midplane_flux}
\end{equation}
required to keep the disk symmetric about the midplane,
\begin{equation}
2\int^{z_{surf}}_{z=0}\rho dz = \Sigma,
\label{eq:dens_converge}
\end{equation}
so that the volume densities add up to the surface density in
the annulus, and
\begin{equation}
\beta(z=0) = 1000.
\label{eq:constant_beta}
\end{equation}
Equation \ref{eq:constant_beta} requires that the plasma $\beta$ be
constant throughout the disk midplane. After turbulence is established,
\citet{FN06} find midplane values of $100 \la \beta \la 1000$ for a wide
range of vertical box sizes, resolutions and boundary conditions in
global ideal MHD calculations. We use the Newton-Raphson algorithm
\citep{Press92} to adjust the initial guesses of $T_{acc}$, $\rho$ and
$P_{B,z}$ until a solution is found that satisfies Equations
\ref{eq:midplane_flux}, \ref{eq:dens_converge}, and
\ref{eq:constant_beta}.

The on-off nature of the MRI creates discontinuities in and $\nu(z)$ and
$F_{acc}(z)$ that can cause the Newton-Raphson algorithm to oscillate
between two sets of input parameters that bracket the correct solution
of Equations \ref{eq:midplane_flux}, \ref{eq:dens_converge} and
\ref{eq:constant_beta}.  To avoid such oscillations in the Ohmic regime,
we decrease the value of $\alpha$ gradually in the range
$1.6\geq\Lambda\geq 0.4$, using a sigmoid function:
\begin{equation}
\alpha\rightarrow\alpha/(1+e^{-k(\Lambda-1)}),
\label{eq:sigmoid}
\end{equation}
where $k = 1.5 \ln (1 / (\alpha_{min}\beta)-1)$.  We use a similar sigmoid
function to smooth $\alpha$ in the ambipolar regime for
$0.4\beta_{min}\leq\beta\leq 1.6\beta_{min}$.

\section{Mass Transport in MRI-Active Disks}
\label{sec:mass_transport}

Here we present our simulations of the evolution of magnetically
turbulent disks over million-year timescales. In \S
\ref{sec:turbulent_structure}, we discuss the relative sizes of the dead
zone and active layers (Question 1 in Introduction). In \S
\ref{sec:mass_flow}, we demonstrate how the shrinking of the active
layer over regions of high density enhances mass pileup in the dead
zone. We also analyze the radial mass flow $\dot{M}(R)$ through the disk
and show that the disk never reaches a steady state, even on
million-year timescales (Question 2 in Introduction). In \S
\ref{sec:prescription}, we give a simple prescription for
accretional heating due to MRI for use with semi-analytical,
non-evolving disk models to predict observables (Question 3 in
Introduction).

\subsection{Turbulent Structure}
\label{sec:turbulent_structure}

Figure \ref{fig:visc30AU} shows viscosity as a function of $(R,z)$ for
the inner 30~AU of Model 1 ($0.015 M_{\odot}$, top) and Model 2 ($0.03
M_{\odot}$, bottom). Outside of 30~AU, viscosity is almost
independent of height $z$.  The left-hand panels of Figure
\ref{fig:visc30AU} show the disks after $10^4$ years of simulation
time---a star age of 0.11~Myr, since we began the simulations with a
star at the beginning of the T-Tauri phase, age 0.1~Myr
\citep{dunham12}. The right-hand panels show both model disks after
1~Myr of simulation time. The plots reveal two important
features of the evolution of low-mass, MRI-active disks:

\begin{figure*}[ht]
\centering
\begin{tabular}{cc}
\includegraphics[trim=3cm 2cm 0cm 11cm, clip=true,
width=0.48\textwidth]{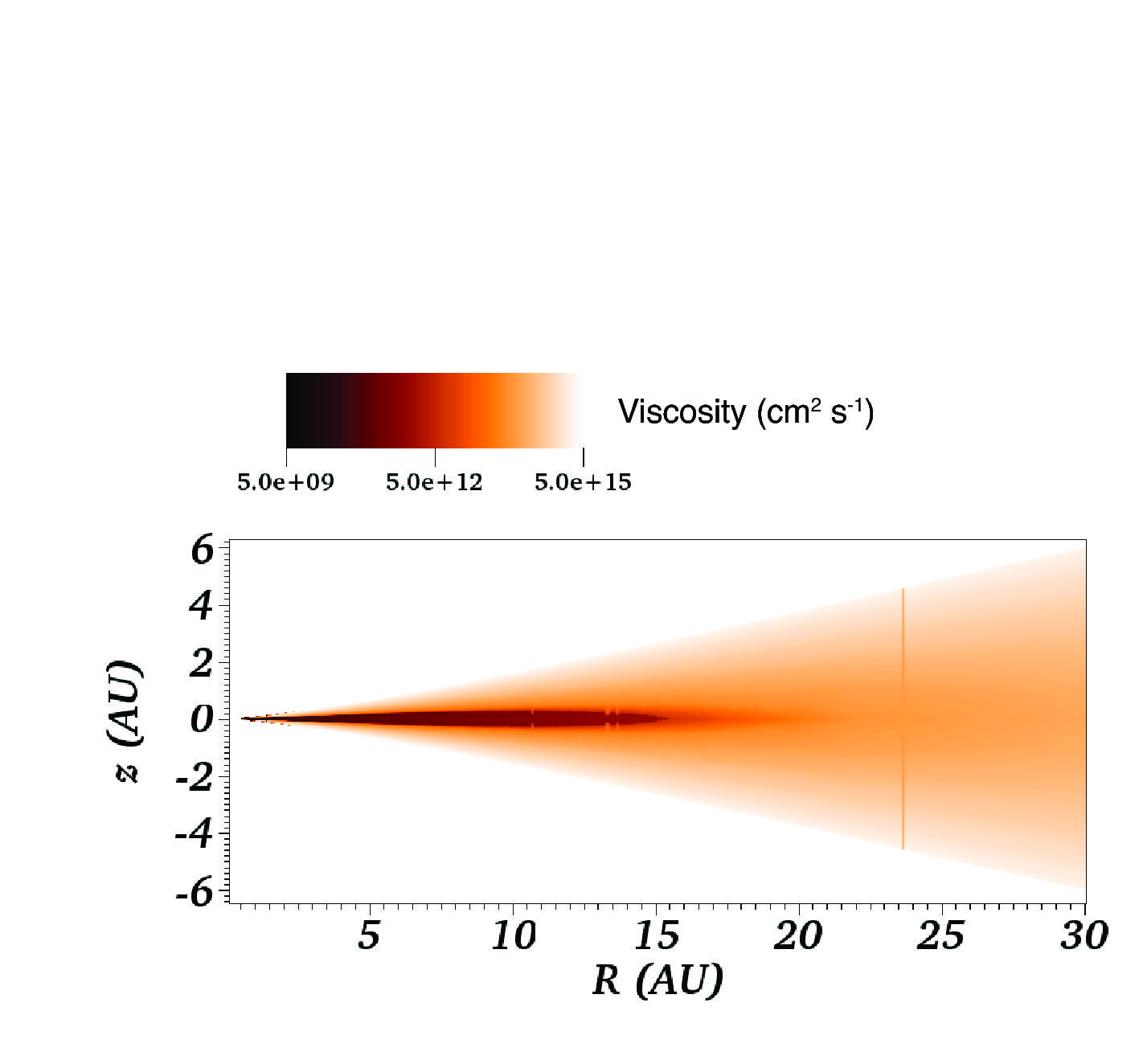}
\includegraphics[trim=3cm 2cm 0cm 11cm, clip=true,
width=0.48\textwidth]{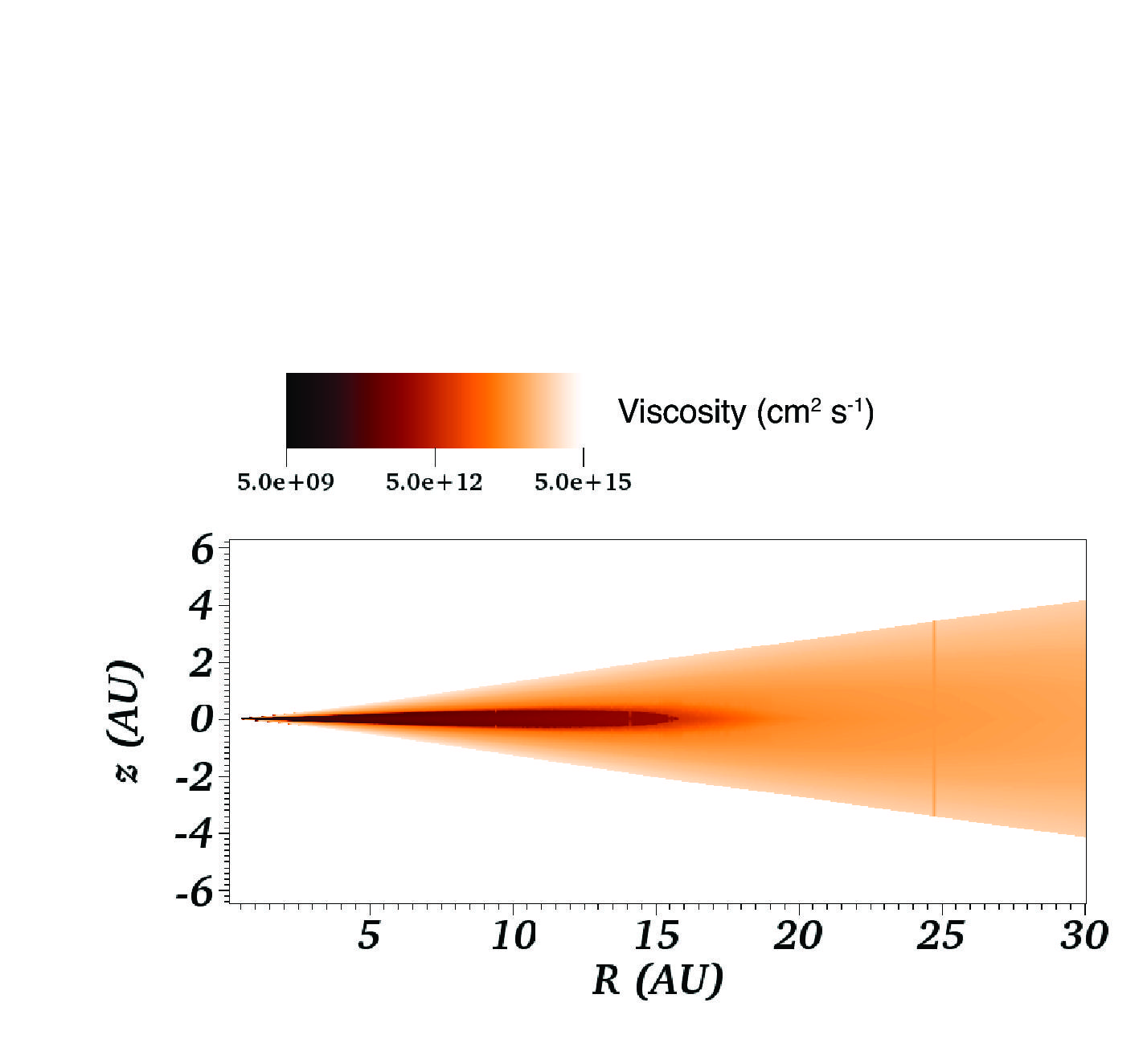} \\
\includegraphics[trim=3cm 2cm 0cm 11cm, clip=true,
width=0.48\textwidth]{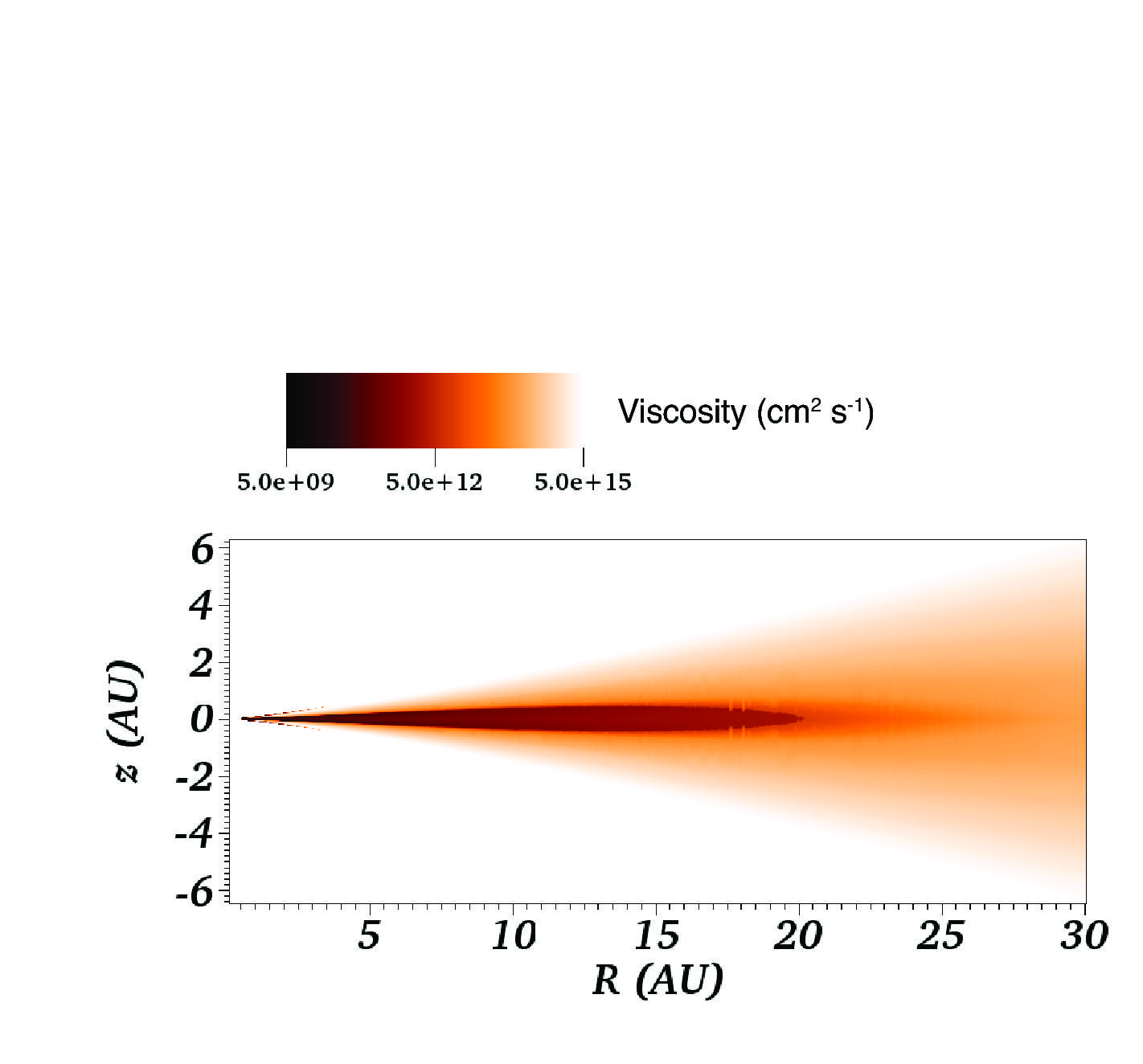}
\includegraphics[trim=3cm 2cm 0cm 11cm, clip=true,
width=0.48\textwidth]{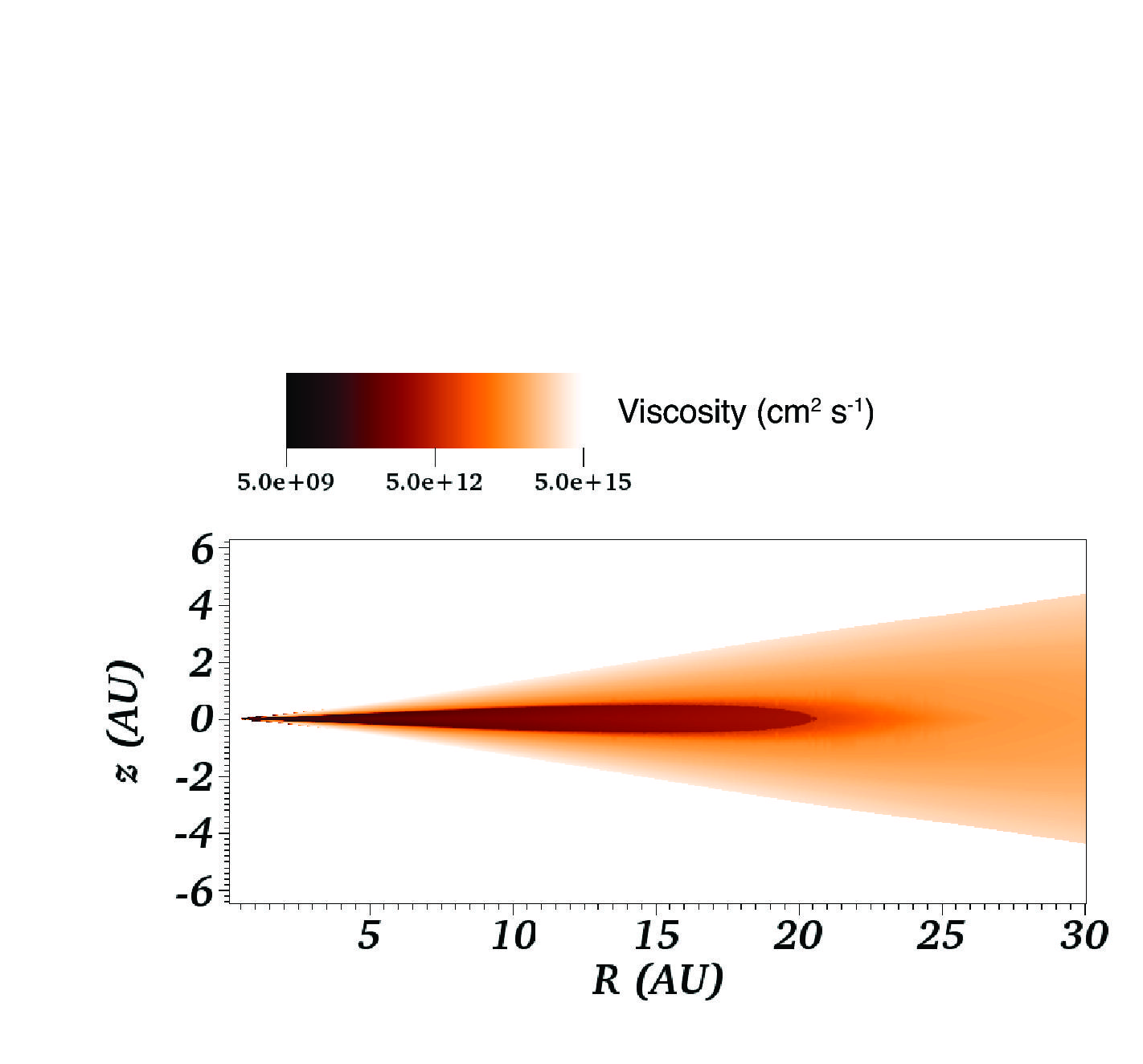}
\end{tabular}
\caption{Viscosity as a function of $(R,z)$ at two
timepoints during disk evolution. The more massive Model 2 disk has the
larger dead zone, which extends past 20~AU. In both models, the radial
extent of the dead zone stays approximately constant with time.
{\bf Top left}: Model 1 (disk mass $0.015
M_{\odot}$), $10^4$ years of simulation time (star age 0.11~Myr). {\bf
Top right}: Model 1 after 1~Myr of simulation time. {\bf Bottom left}:
Model 2 after $10^4$ years of simulation time. {\bf Bottom right}: Model
2 after 1~Myr of simulation time.}
\label{fig:visc30AU}
\end{figure*}

\begin{enumerate}

\item A midplane dead zone, where Ohmic diffusion quenches the MRI and
restricts viscosity, extends to 16~AU in Model~1 and 21~AU in
Model~2.

\item Although the vertical heights of both the dead zone and the
overall disk shrink with time as the disk loses mass and cools (see
Section \ref{sec:thermo}), the radial extent of the dead zone stays
approximately constant in time.

\end{enumerate}

The radial size of the dead zone is larger than the $\sim 5$~AU
typically quoted for disks similar to the minimum-mass solar nebula
(MMSN) \citep{matsumura03, salmeron08, flaig12}. Here we define the dead
zone as the region of the disk where $\Lambda < 1$.  The primary reason
our dead zone is so extensive is the surface density of our disks: the
MMSN contains $0.01 M_{\odot}$ within 100~AU of the sun, while our least
massive disk contains $0.015 M_{\odot}$ within 70~AU. Our dead zone is
also extensive because we treat ambipolar diffusion, as well as Ohmic
diffusion. \citet{sano00} use the same tenfold dust depletion we do,
but treat only Ohmic diffusion and find a dead zone that extends
5--10~AU.

Dead zones with $\alpha \simeq 10^{-5}$ are required for maintaining
compositional gradients such as the ice gradient in the asteroid belt,
which would radially diffuse on million-year timescales for fully
turbulent disks \citep{nelson10}. The fact that Uranus and Neptune have
atmospheres with higher carbon enrichment than Jupiter and Saturn
provides some evidence that the solar nebula had compositional gradients
covering the entire giant planet-formation region \citep{encrenaz05}.
Given that the dead zone in a disk of just $0.015 M_{\odot}$ can reach
$\sim 15$~AU, the outer boundary of the giant planet-formation region in
the Nice model \citep{tsiganis05}, the solar nebula could have supported
such a compositional gradient. The lack of change in the dead zone
radius over time suggests that compositional gradients are stable over
million-year timescales.

\begin{figure*}[ht]
\centering
\begin{tabular}{c}
\includegraphics[trim=2.5cm 2cm 0cm 12.5cm, clip=true,
width=0.65\textwidth]{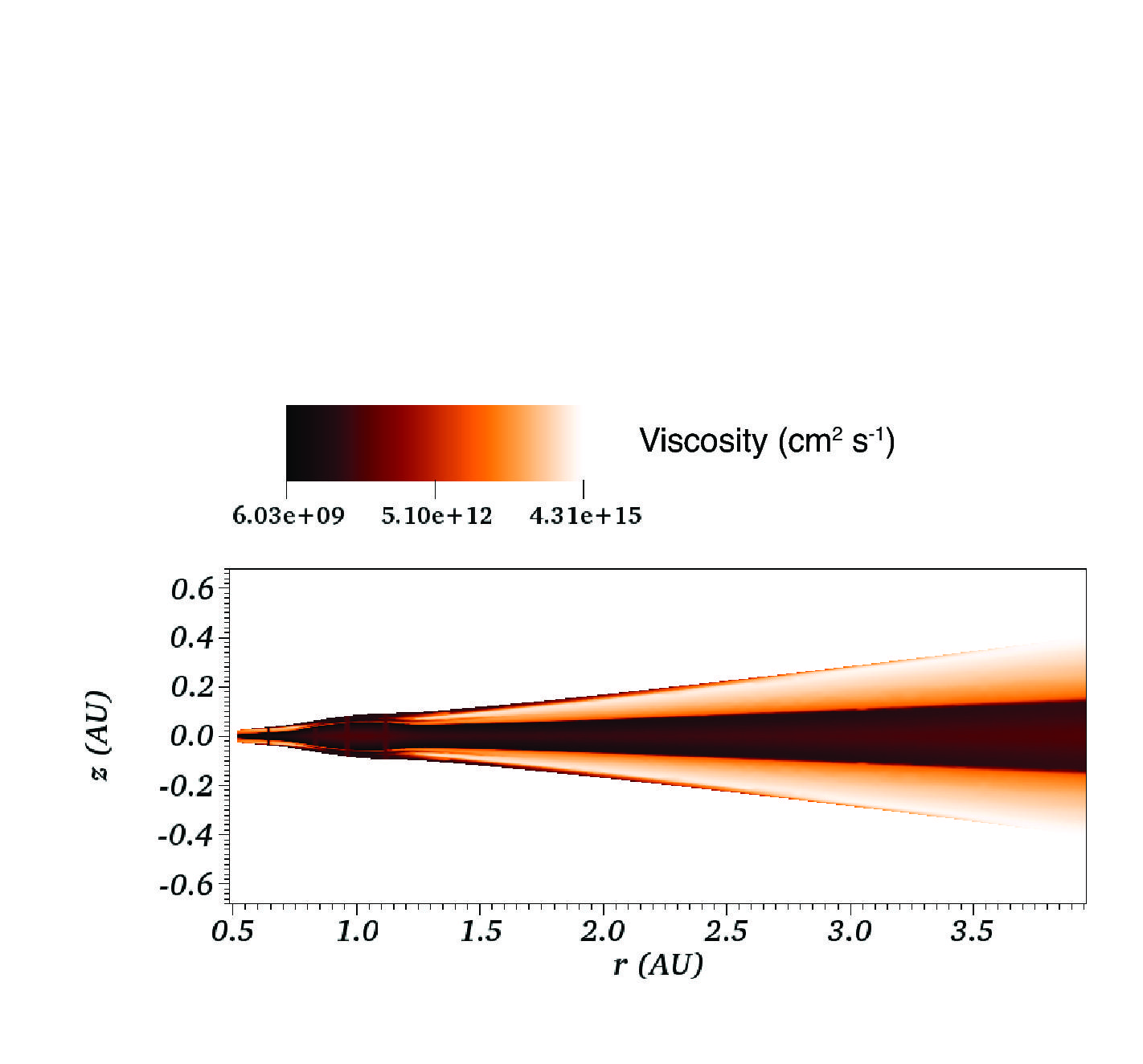} \\
\includegraphics[trim=2.5cm 2cm 0cm 12.5cm, clip=true,
width=0.65\textwidth]{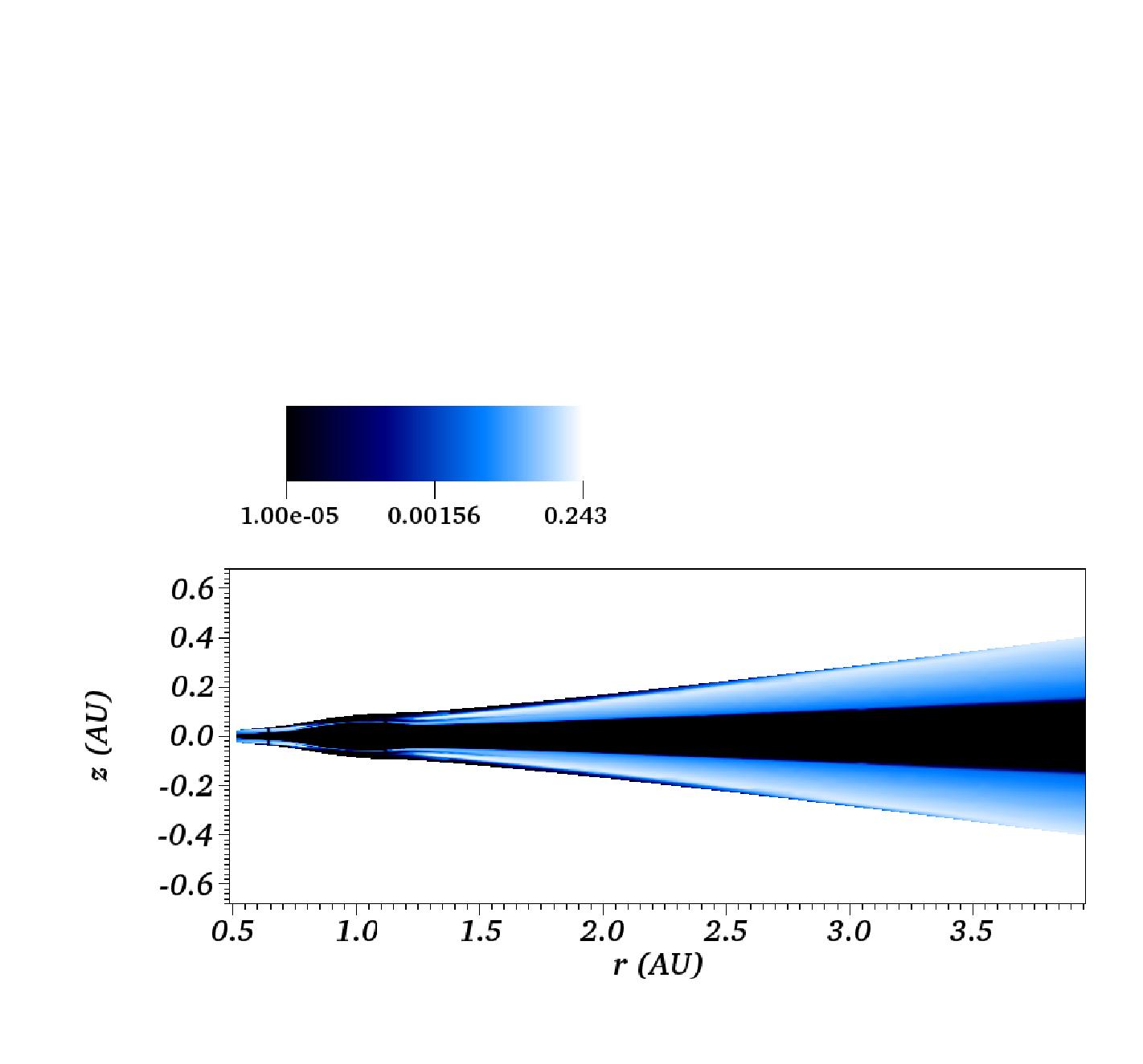} \\
\includegraphics[trim=2.5cm 2cm 0cm 12.5cm, clip=true,
width=0.65\textwidth]{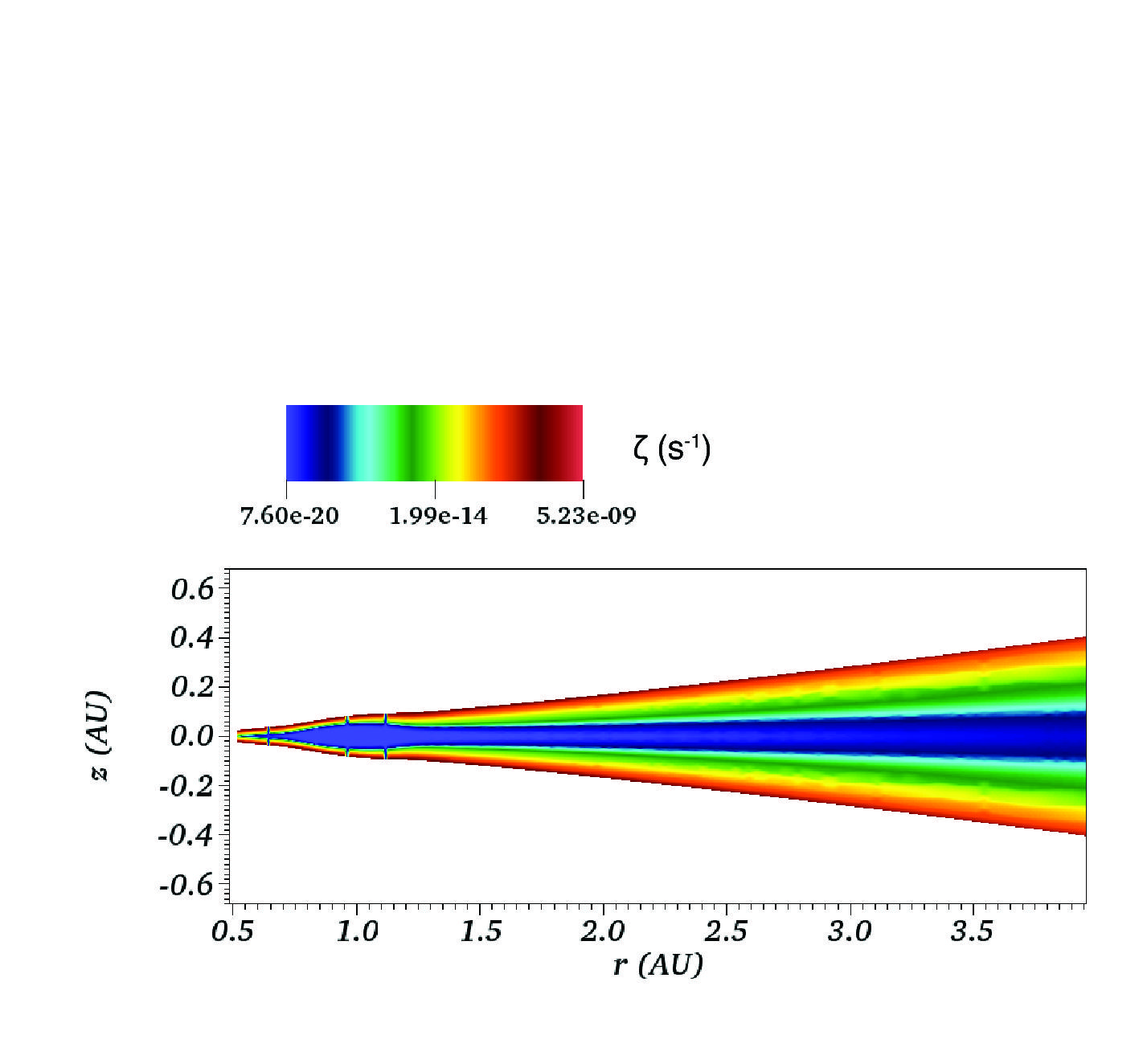}
\end{tabular}
\caption{Plots of viscosity $\nu$ ({\bf top}),
turbulent stress coefficient $\alpha$ ({\bf middle}), and ionization
rate $\zeta$ ({\bf bottom}) for the inner 4~AU of Model 2 at
1~Myr. In the inner 3~AU, a stable corona where ambipolar diffusion
shuts down MRI sits on top of the active layer. Note the thinness of the
active layer over the pileup at 1~AU. The dead slice seen in
Figure \ref{fig:vstructline} is also visible between 1.2~AU and 1.7~AU.}
\label{fig:vstructzoom}
\end{figure*}

Figure \ref{fig:vstructline} gives further insight into the vertical
structures of Model 1 (top) and Model 2 (bottom). In each panel, the
active layer is shaded with red dots. At 10~AU, both disks show the
classical layered structure of an MRI-active zone on tope of a dead
zone. At 1.5~AU, however, we see additional complexity in the disk's
vertical structure. At 45~mG (Model 1) and 70~mG (Model 2), the magnetic
tension is strong enough to prevent MRI from bending the field lines in
the low-density surface gas despite a high ionization rate of
$10^{-10}$~s$^{-1}$, forming a stable corona \citep[e.g.][]{MS2000}.
Our stabilizing magnetic field values are roughly consistent with those
of \citet{salmeron08}, who calculated that unstable MRI modes can only
grow for $B \la 80$~mG in the presence of $1 \mu m$ grains. (See \S
\ref{subsec:regime} and Table \ref{table:diskpars} for more on the grain
properties used in our models.) Figure \ref{fig:vstructzoom} is a
zoom-in on the stress coefficient $\alpha(R,z)$, viscosity $\nu(R,z)$
and ionization rate $\zeta(R,z)$ in the inner 4~AU of Model 2 after
1~Myr. The inactive corona is apparent for radii $R \la 3$~AU, but moves
above our computed disk surface (the location where $\tau_*^{\|} \sim
1$) as the magnetic field stength decreases with radius.

Also noticeable in Figures \ref{fig:vstructline} and
\ref{fig:vstructzoom} is a split active layer in the inner part of Model
2. MRI-active regions of high $\alpha$ and $\nu$ sandwich a ``dead
slice'' that has reduced stress and viscosity by an order of magnitude.
(The fact that $\alpha$ does not immediately plunge to its minimum value
in the dead slice is a result of the sigmoid smoothing described in \S
\ref{subsec:methods}.) While not present at the start of our
simulations, the split active layer appears after only 5000 years of
disk evolution and persists until the end of the simulation at 3~Myr.
The split active layer extends from roughly $1.2$~AU~$< R < 1.7$~AU
(Figure \ref{fig:vstructzoom}), though its radial extent shrinks slightly
as the disk evolves.

The dead slice, in the part of the disk where ambipolar diffusion is the
strongest non-ideal MHD term, is the result of two competing effects.
First, MRI in the ambipolar regime requires high ionization: decreasing
$\zeta$ toward the disk midplane reduces $Am$ and shuts down turbulence.
Yet ambipolar diffusion is quenched at high densities: increasing $\rho$
toward the midplane increases $Am$, favoring turbulence. In the dead
slice, the dropping $\zeta$ temporarily dominates over the increasing
$\rho$ and shuts down the MRI. Model 1 does not have a dead slice
because its surface density is about 1/2 that of Model 2: $\zeta$ can
stay high enough for the MRI to operate until very near the midplane,
where Ohmic diffusion begins to dominate. An open question is whether or
not a dead slice would be present in a fully 3-D simulation with
identical vertical profiles of $\rho$, $\zeta$ and $\beta$ to our disk:
the thickness of the dead slice is smaller than the MRI wavelength by a
factor of 2--10, so turbulence would very likely erase it.

We have seen that MRI-active disks may have complex vertical
structure, with multiple layers of turbulent and non-turbulent
zones. In the next section we explore the overall mass flow
through the disk and discuss its evolution on million-year
timescales.

\subsection{Mass Flow}
\label{sec:mass_flow}

\begin{figure*}[ht]
\plottwo{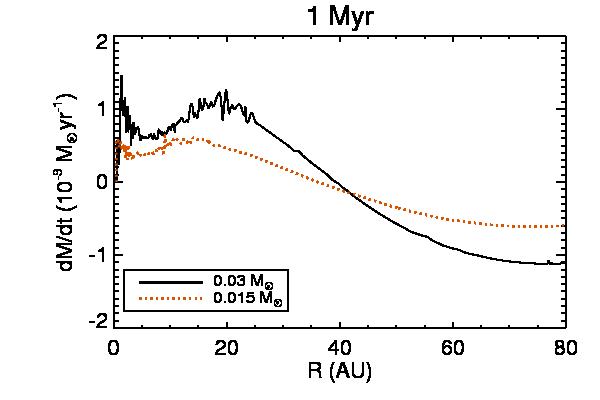}{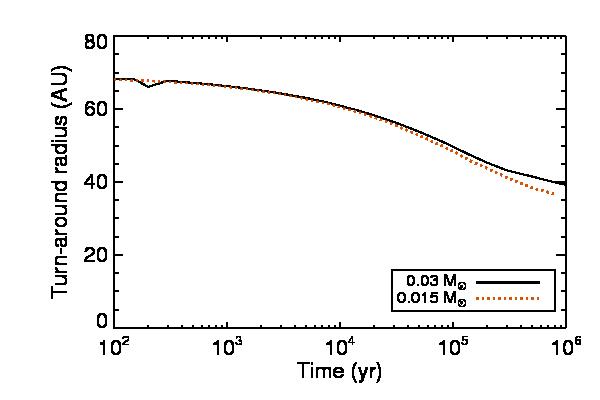}
\caption{{\bf Left}: A plot of $\dot{M}(R)$ for both disk models after
1~Myr of disk evolution reveals that low-mass MRI-driven
accretion disks have not reached steady state after 1~Myr. As required by
conservation of angular momentum, there is a turn-around in the
mass flow between inward accretion and outward decretion. The
highest accretion rates in the inner disk, where mass moves
toward the star, occur on either side of the dead zone. {\bf
Right}: The location of the turn-around radius as a function of
time for both models. The turn-around radius moves inward over
time.}
\label{fig:mdot}
\end{figure*}

As might be expected for a layered disk with a dead zone, radial mass
transport is not in steady state for either Model 1 or Model 2.  The
left-hand panel of Figure \ref{fig:mdot} shows \begin{equation}
\dot{M}(R) = 2 \pi R \Sigma v_R \label{eq:mdot} \end{equation} for both
models after 1~Myr of evolution, where $v_R$ is a density-weighted,
vertically averaged gas radial velocity.  The convention in Equation
\ref{eq:mdot} is that $v_R$ is positive when gas flows toward the star
and negative when gas flows away from the star. In the inner part of the
disk, where gas flows inward, the highest accretion rates occur (a) at
the outer edge of the dead zone (at 21~AU for Model 1 and 16~AU for
Model 2), and (b) at the inner disk boundary.  There is a clear drop in
$\dot{M}(R)$ associated with the dead zone. (The noisy $\dot{M}$ profile
where the disk has two or more layers is a reflection of the
root-finding tolerance in our Newton-Raphson algorithm.  Fluctuations in
the location of the boundary between dead and active zones are random
and average out over time.)

\begin{figure*}[ht]
\centering
\begin{tabular}{cc}
\includegraphics[trim=2cm 2cm 0cm 12cm, clip=true,
width=0.48\textwidth]{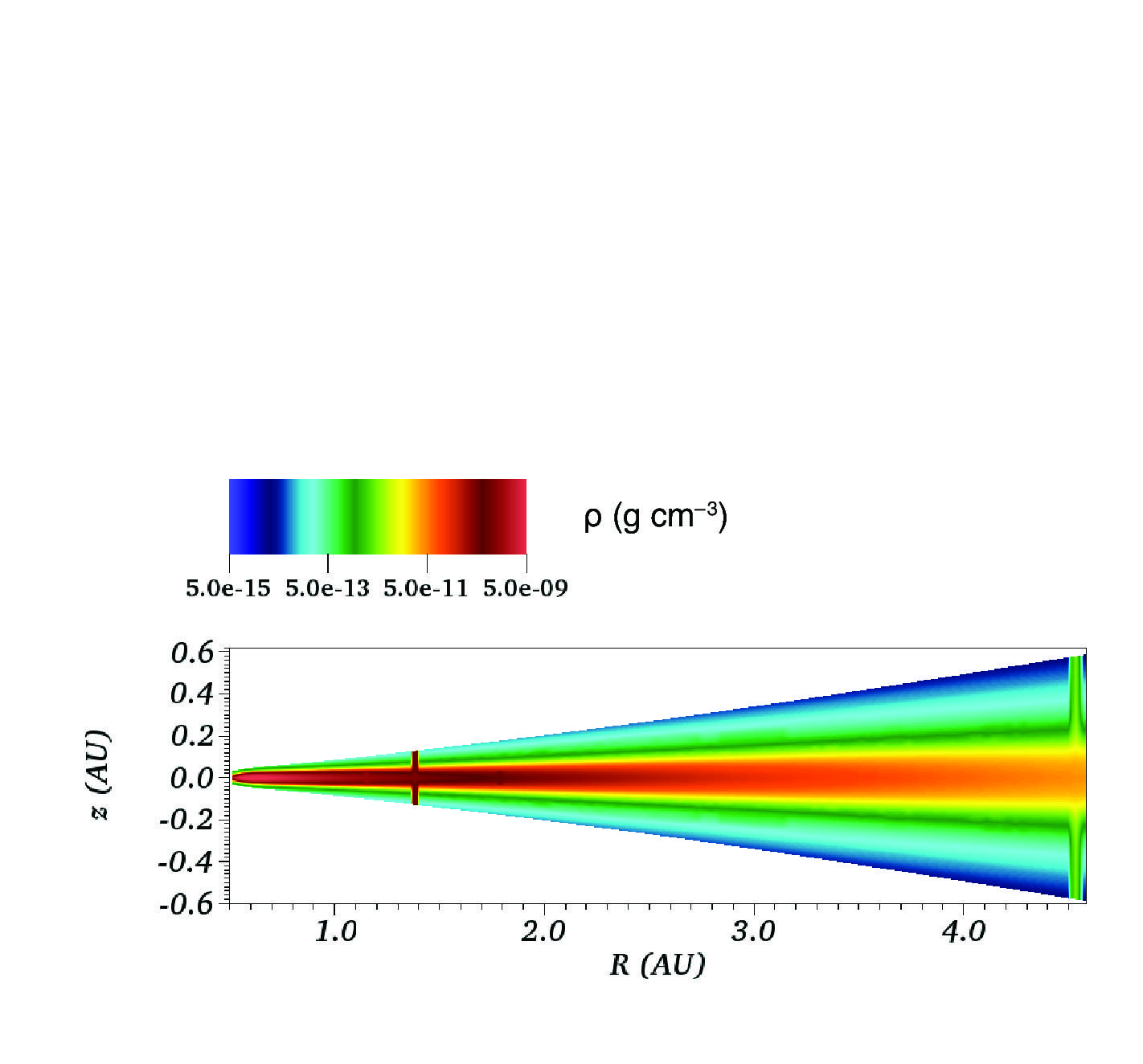}
\includegraphics[trim=2cm 2cm 0cm 12cm, clip=true,
width=0.48\textwidth]{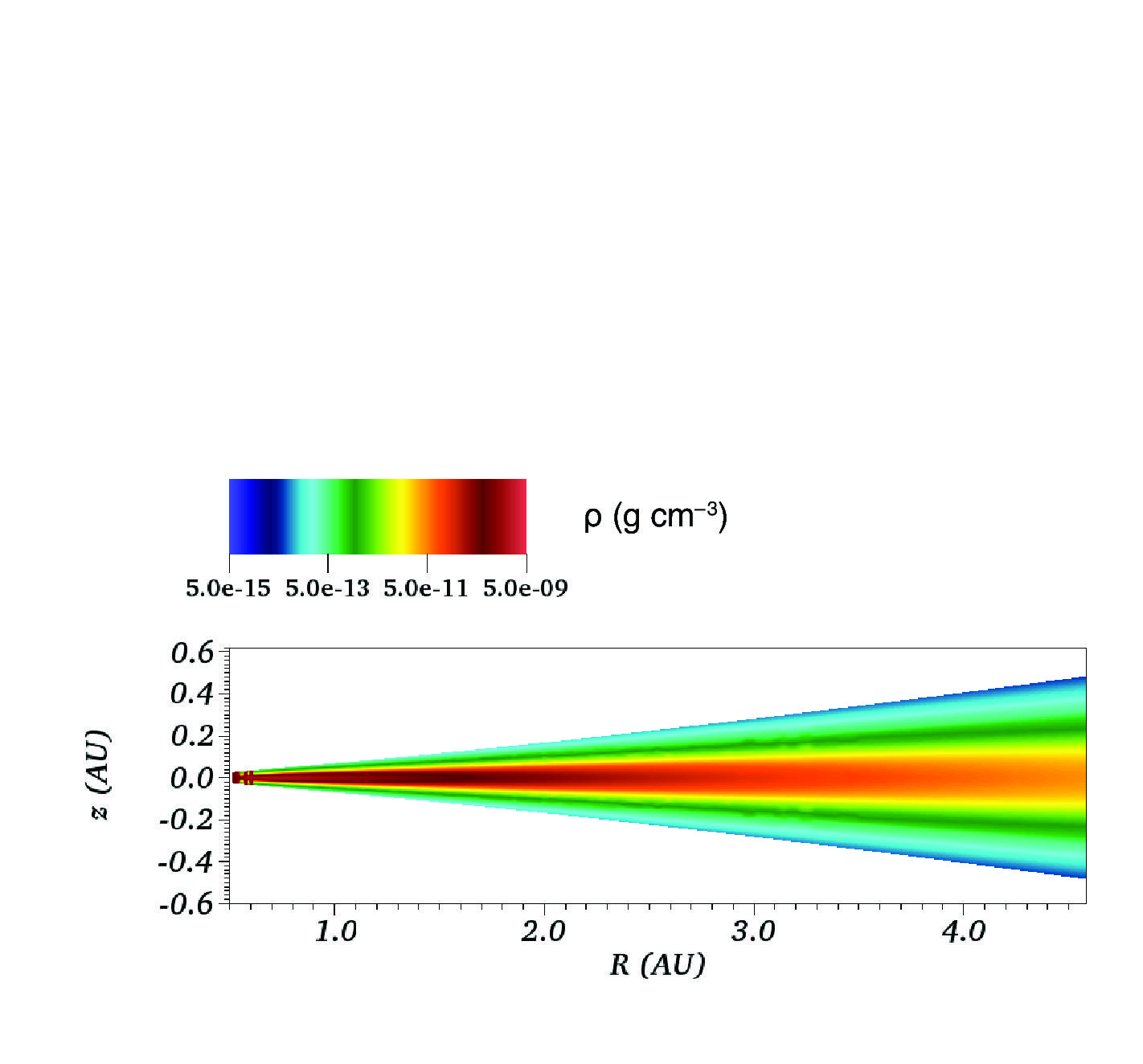} \\
\includegraphics[trim=2cm 2cm 0cm 12cm, clip=true,
width=0.48\textwidth]{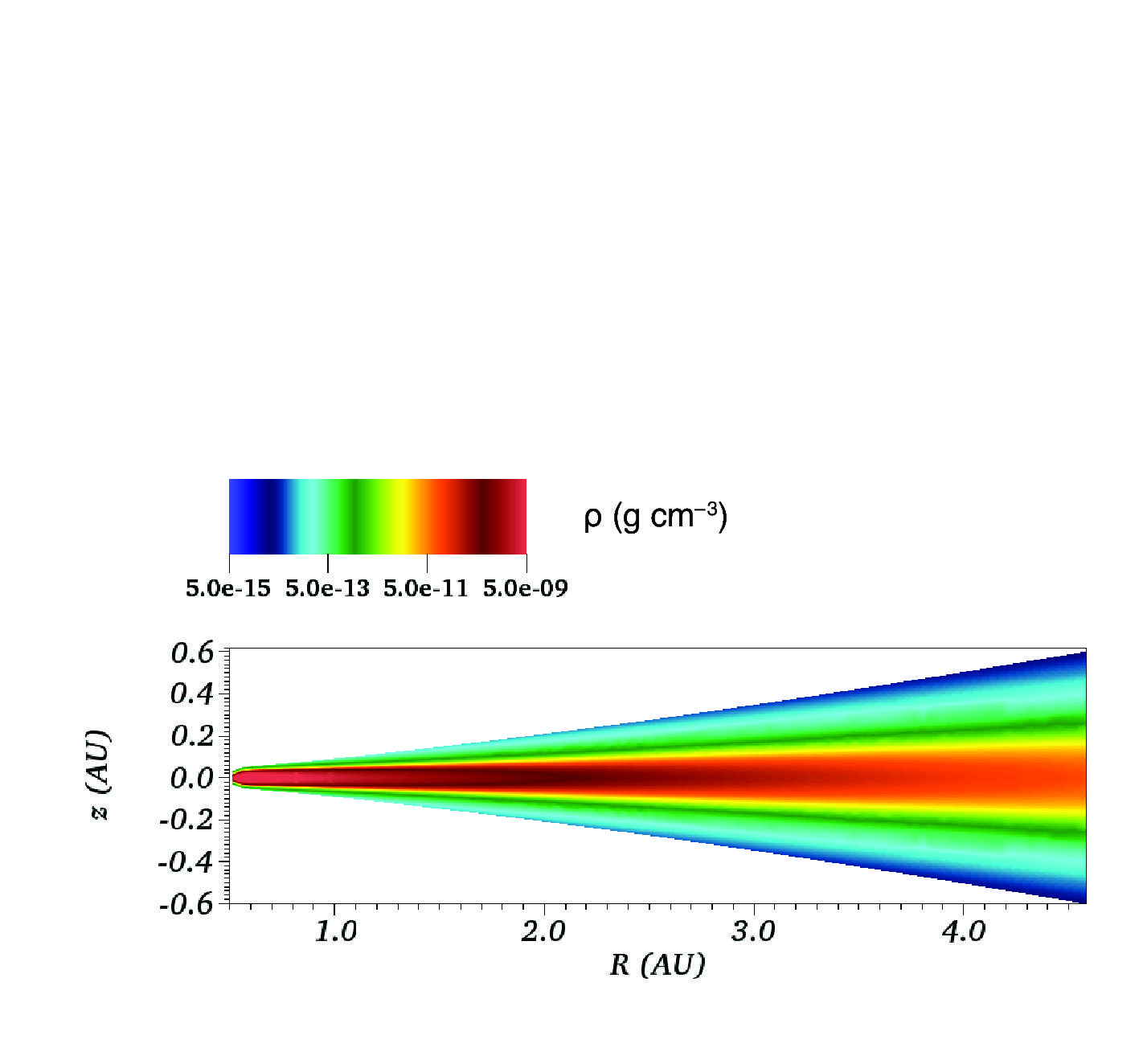}
\includegraphics[trim=2cm 2cm 0cm 12cm, clip=true,
width=0.48\textwidth]{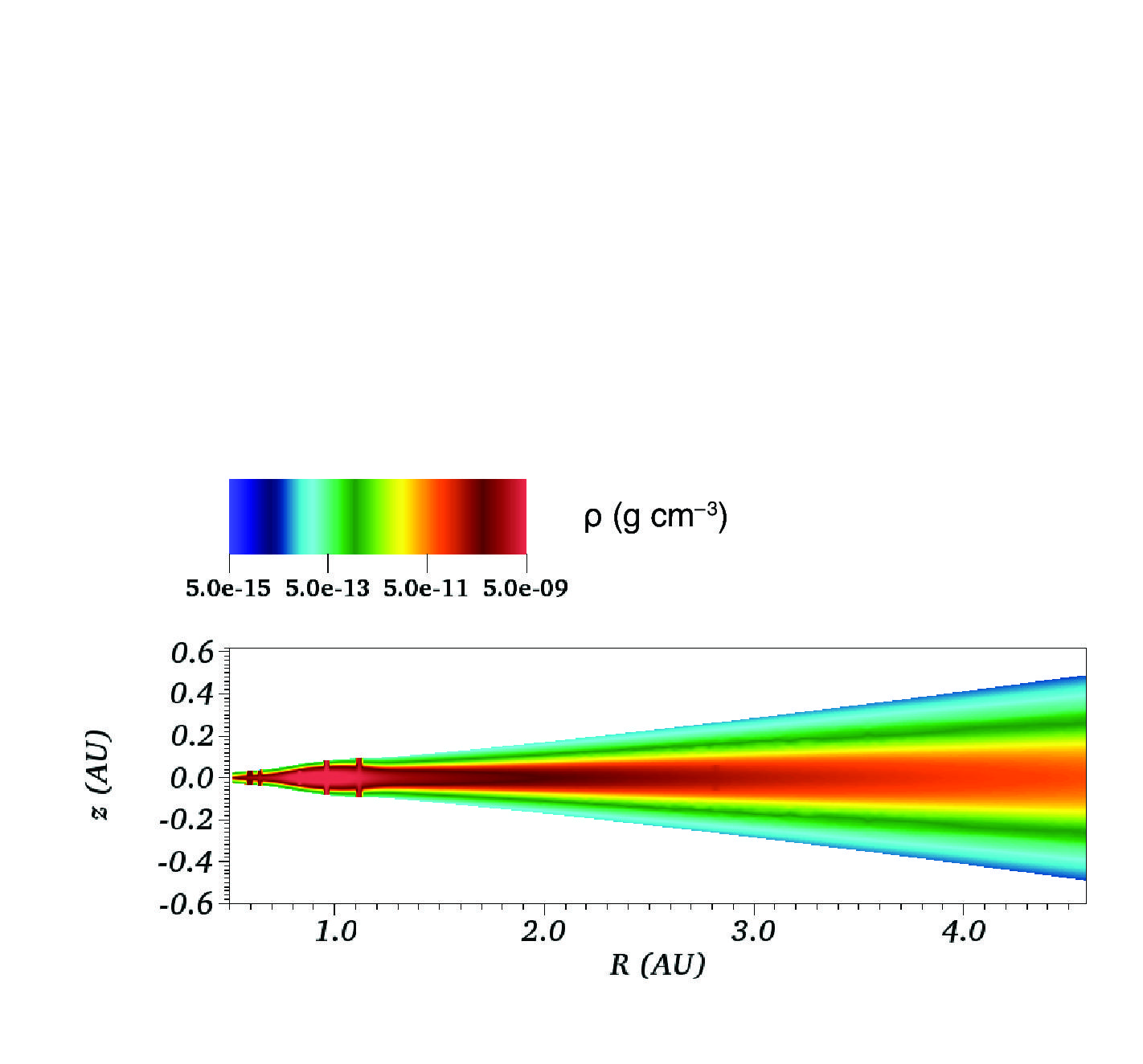}
\end{tabular}
\caption{Volume density as a function of $(R,z)$ at two
timepoints during disk evolution. In the more massive Model 2, the
density in the pileup at 1~AU noticeably grows with time.  {\bf Top
left}: Model 1 (disk mass $0.015 M_{\odot}$), $10^4$ years of simulation
time (star age 0.11~Myr). {\bf Top right}: Model 1 after 1~Myr of
simulation time. {\bf Bottom left}: Model 2 after $10^4$ years of
simulation time. {\bf Bottom right}: Model 2 after 1~Myr of simulation
time.}
\label{fig:dens4AU}
\end{figure*}

The potential for the dead zone to become gravitationally unstable,
given enough time to accumulate mass, is clear \citep{Gammie96,
armitage01, zhu10}---though we see only a slow, steady density growth at
$\sim 1$~AU--4~AU over the course of 3~Myr in Model 2 (Figure
\ref{fig:dens4AU}). We will examine the potential for gravitational
instability in high-mass, MRI-active disks in a forthcoming paper. One
important caveat is that the accumulation of mass in the dead zone is
unstable to the Rossby wave instability \citep{hawley87, li01,
meheut12}, which triggers spiral density waves. Without azimuthal
information in our model, we cannot model the effect of the Rossby wave
instability on dead-zone overdensity.  Ultimately, the overdensity may
not survive and may break up into large-scale vortices
\citep{papaloizou85, lovelace99}.

At $t = 1$~Myr, there is turn-around in the accretion flow near 40~AU in
each model. The outward mass flow outside 40~AU is necessary for overall
conservation of angular momentum as material in the inner disk moves
toward the star. The fact that there must be a change in the accretion
flow direction also follows directly from the diffusive nature of
T-Tauri disks \citep[e.g.][]{lyndenbell74}, which lack a circumstellar
envelope to feed steady-state inward accretion. In both models, the disk
has expanded beyond its initial 70~AU size after 1~Myr. Note that
the 40~AU region of the disk does not drain, as the ``turn-around
radius'' where the accretion flow changes direction moves inward with
time (right-hand panel of Figure \ref{fig:mdot}). We heartily discourage
the use of a steady-state accretion rate for any reasonable description
of angular momentum transport in T-Tauri disks.  However, for predicting
observables to order-of-magnitude using a static, non-time-varying disk 
model, we suggest approximating the accretional heating in MRI-active 
disks with a constant $\dot{M} \sim 10^{-9} M_{\odot}$~yr$^{-1}$.
Figure \ref{fig:mdot} suggests that at a given time, $\dot{M}$ has a 
modest dependence on disk mass.

How consistent are our modeled accretion rates with observations? An
oft-quoted value of $\dot{M}_*$, the T-Tauri star accretion rate, is
$10^{-8} M_{\odot}$~yr$^{-1}$ \citep{Sicilia04}. Accretion rates in
transitional disks are roughly $3 \times 10^{-9} M_{\odot}$~yr$^{-1}$
\citep{espaillat12}. Our simulations achieve an accretion rate of
roughly $\dot{M}_* \sim 10^{-9} M_{\odot}$~yr$^{-1}$, more consistent
with the median transitional disk accretion rate. Reproducing high
accretion rates in the inner regions of MRI-active disks is a delicate
balance between magnetic field strength, X-ray ionization, grain size,
and gas/small grain mass ratio. For example, we performed simulations
using the same disk masses and X-ray luminosity (0.015$M_{\odot}$ and
0.03~$M_{\odot}$; $2 \times 10^{30}$~erg~s$^{-1}$, respectively) but a
grain size of $0.1 \mu$m instead of $1 \mu$m and a standard gas/small
grain mass ratio of 100, and quenched all MRI activity in the disk
entirely. However, as models by \citet{zsom11} show grain growth and
settling within 1000~years, it is reasonable to assume the gas/small
grain mass ratio has evolved from the interstellar value by the T-Tauri
phase. In our simulations, a stronger magnetic field (lower value of
$\beta$ at the midplane) also suppressed MRI-driven accretion, which
occurs in regions of the disk dominated by ambipolar diffusion.  Most
simulations of non-ideal MRI-driven accretion have difficulty reaching
the $10^{-8} M_{\odot}$~yr$^{-1}$ benchmark \citep[e.g.][]{zhu10,
Bai11}, though there is a substantial amount of scatter in both T-Tauri
and transitional disk accretion rates \citep{romero12}.




\subsection{Simple prescription for accretional heating
in an MRI-active disk}
\label{sec:prescription}

Previous studies have often relied on
static, non-evolving disk models to connect
measured line fluxes, velocities or spectral indices to physical
properties of disks \citep[e.g.][]{Dalessio06, pinte10}. Unfortunately
the constant-$\alpha$ prescription for turbulent angular momentum
transport leads to unphysical thermodynamic descriptions of disk
midplanes. Predicted midplane temperatures in a constant-$\alpha$ model,
where most turbulence is concentrated at the midplane, are too high.
The poor match of the constant-$\alpha$ model to realistic protostellar
disk accretion is definitely a problem for observations that trace disk
midplanes, such as sub-millimeter measurements of continuum emission
from large grains or surveys of rare molecules like HCO$^+$ or HD.
However, the problem also affects observations that trace surface
layers, such as {\it Spitzer} emission lines, infrared SEDs and
sub-millimeter maps of abundant gases like CO. If a significant subset
of T-Tauri disks rely on MRI to drive accretion, overestimating their
midplane temperatures leads to overpredicting the photosphere and
pressure scale heights, underpredicting the optical depths of
low-excitation lines such as CO ($J = 2 \rightarrow 1$), and
possibly underpredicting the rates of grain settling and growth.

\begin{figure*}[ht]
\plottwo{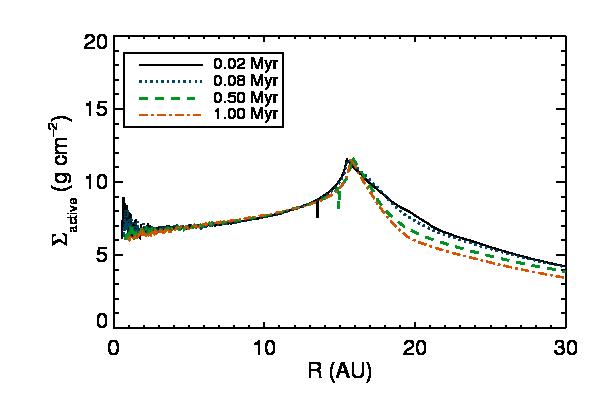}{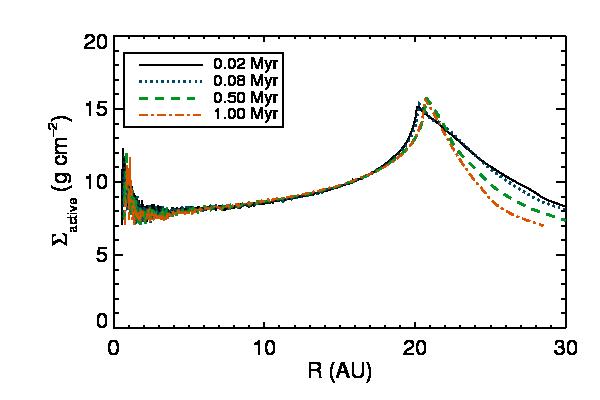}
\caption{$\Sigma_{\rm active}(R)$ for four time
snapshots of Model 1 ({\bf left}) and Model 2 ({\bf right}).
The active column depth as a function of radius changes very
little as the disk evolves, making it possible to approximate
$\Sigma_{\rm active}(R)$ for use in a semi-analytical viscosity
prescription.}
\label{fig:activedepth}
\end{figure*}

Here we use our simulation results to present a simple prescription for
$\alpha$ in an MRI-active disk. First we determine the depth of the
MRI-active layer. Figure \ref{fig:activedepth} shows the surface density
of the active layer as a function of radius for four time snapshots of
Model 1 (left) and Model 2 (right). Clearly, the functional form of
$\Sigma_{\rm active}(R)$ is the same for both models and does not vary
significantly with time. Near the star, where X-ray ionization is
important, $\Sigma_{\rm active}$ is high but falls off quickly as X-ray
irradiation declines (see Equation \ref{eq:xray}). Over the dead zone,
in the region where only cosmic-ray ionization is important,
$\Sigma_{\rm active}(R)$ increases as the magnetic field weakens,
shrinking the inactive corona. $\Sigma_{\rm active}(R)$ reaches its
maximum at the outer edge of the dead zone.  Here, where the entire
vertical column is active, the decreasing depth of the active layer
simply reflects the fact that $\Sigma(R)$ is a decreasing function.

Although $\Sigma_{\rm active}$ does vary with $R$ in the parts of the
disk with a dead zone, the variation is less than a factor of two for
both models presented here. Similarly, $\Sigma_{\rm active}(R)$ is
slightly higher for Model 2 than for Model 1, suggesting that the active
column may increase modestly with disk mass. $\Sigma_{\rm active}
\approx 10$~g~cm$^{-2}$ is a good approximation for low-mass disks with
surface density less than about four times the MMSN. The semi-analytical
viscosity prescription is then simple:

\begin{enumerate}

\item Set the depth of the active column to $\Sigma_{\rm active} \approx
10$~g~cm$^{-2}$.  Set $\alpha \approx 0.01$ in the active column and
$\alpha \approx 10^{-5}$ in the dead zone.  While there may be a corona
on the inner disk surface, it contains very little mass and may safely
be ignored as long as the disk model includes stellar heating.

\item Smooth the transition between the active layer and the dead zone
if desired.

\item The entire vertical column will be active where $\Sigma \approx
20$~g~cm$^{-2}$, such that the two active layers meet at the midplane.
Outside the outer radius of the dead zone, where $\Sigma \approx
20$~g~cm$^{-2}$, use $\alpha \approx 0.01$.

\item Calculate viscosity $\nu$ at each $(R,z)$ in the disk
using Equation \ref{eq:alpha}.

\end{enumerate}

\section{Thermal Evolution of MRI-Active Disks}
\label{sec:thermo}

Now that we understand how mass moves through a disk where accretion is
driven by MRI, we turn our attention to the thermal structure and
evolution of the disk. Here we see some important differences from disk
models that use a constant-$\alpha$ viscosity prescription. In Section
\ref{sec:heatcool} we discuss how different parts of MRI-active disks
heat up and cool off over time (Question 4 in Introduction). In Section
\ref{sec:iceline} we investigate the location of the ice line and its
motion through an MRI-active disk (Question 5 in Introduction).

\subsection{Disk Heating and Cooling}
\label{sec:heatcool}

\begin{figure*}[ht]
\centering
\begin{tabular}{cc}
\includegraphics[width=0.48\textwidth]{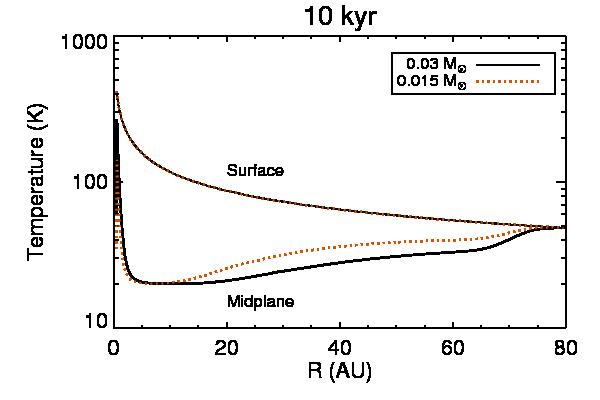}
\includegraphics[width=0.48\textwidth]{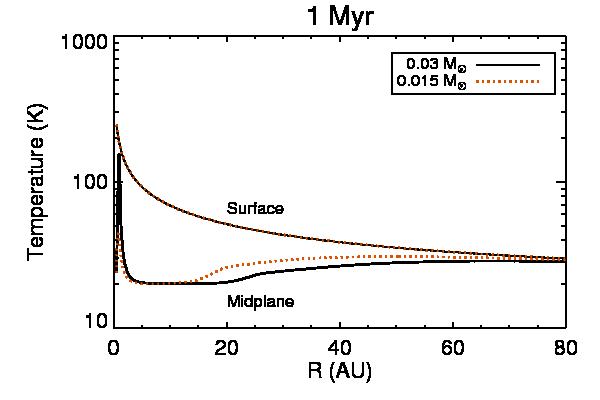}
\\
\includegraphics[trim=1cm 2cm 0cm 12cm, clip=true, width=0.48\textwidth]{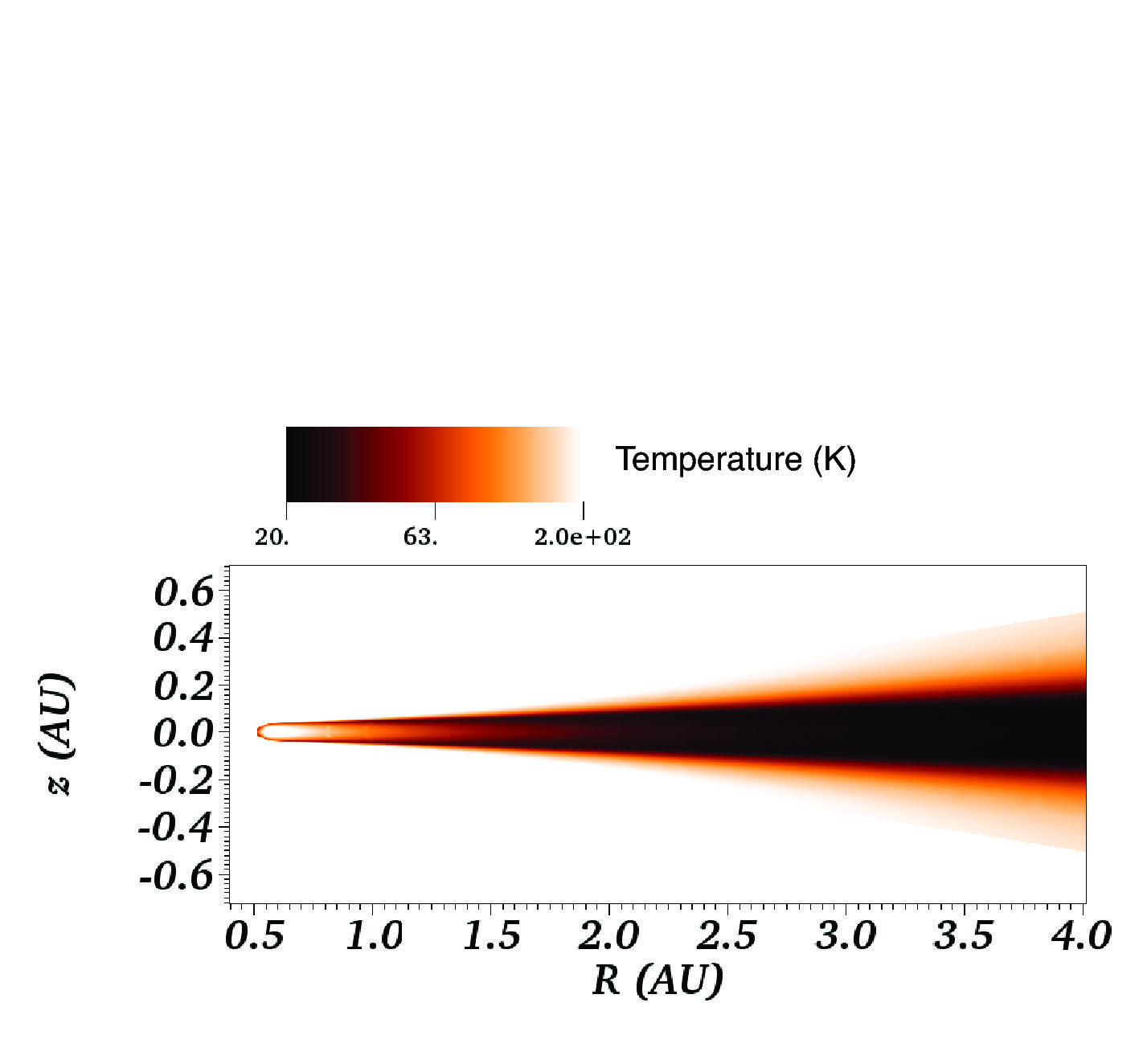}
\includegraphics[trim=1cm 2cm 0cm 12cm, clip=true, width=0.48\textwidth]{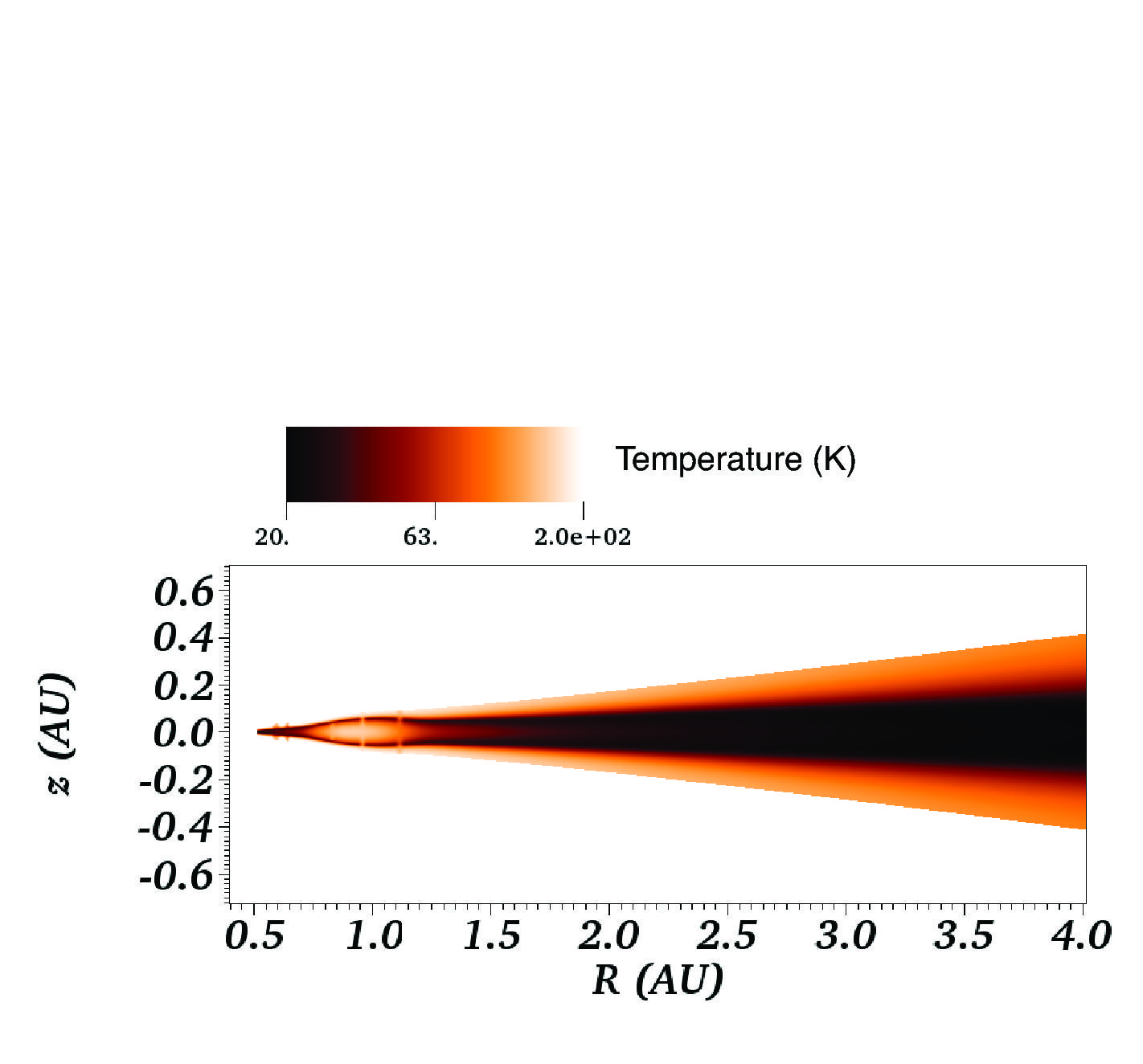}
\end{tabular}
\caption{Although the surface of the disk cools over time, as the
T-Tauri star moves down the Hayashi trach, parts of the midplane of an
MRI-active disk may heat up over time as mass piles up in the dead zone.
{\bf Top left}: Temperature at the surface (top curves) and the midplane
(bottom curves) of Model 1 (red dotted) and Model 2 (black solid) after
10,000~yr. {\bf Top right}: Surface and midplane temperatures of Models 1
and 2 after 1~Myr. Note the increase in midplane temperature at the edge
of the dead zone in each model. {\bf Bottom left}: Contour plot of
$T(R,z)$ in the inner 4~AU of Model 2 after 10,000~yr of evolution.
Temperature units are Kelvins. {\bf
Bottom right}: Contour plot of $T(R,z)$ in the inner 4~AU of Model 2
after 1~Myr of evolution. Note how the pileup of material centered at
1~AU (see Figure \ref{fig:dens4AU}) has heated up over time.}
\label{fig:temperature}
\end{figure*}

The top panels of Figure \ref{fig:temperature} show the surface and
midplane temperatures of Models 1 and 2 after 10 thousand years of 
evolution (left) and 1~Myr of evolution (right). First, note the 
obvious feature that the surface is far hotter than the midplane 
throughout most of the disk.
Disk models with constant-$\alpha$ viscosity prescriptions usually have
warm midplanes, $T \ga 100$~K, in the inner 5~AU. In the inner 1~AU of
constant-$\alpha$ models, the midplane temperature can even exceed the
surface temperature, approaching 1000~K \citep[e.g.][]{Hersant01,
Dalessio06, DR09}. In our models, there is so little turbulent energy
generated in the dead zone that the disk midplane falls to 20~K---the
ambient temperature of the remnant molecular cloud surrounding the disk.
Here we assume the disk is optically thin to long-wavelength radiation
from the ambient cloud and cannot cool below the ambient temperature.
Only in the inner $\sim 2$~AU does residual mass transfer by large-scale
magnetic fields lift the midplane temperature above the minimum value.

Moving outward through the disk, the midplane temperature rises modestly
until it equalizes with the surface temperature where the disk becomes
optically thin to stellar radiation. Both of our models, despite their
modest masses ($0.015 M_{\odot}$ and $0.03 M_{\odot}$), are optically
thick to stellar radiation inside $\sim 70$~AU, decoupling the surface
and midplane temperatures. Passively heated disk models
\citep[e.g.][]{woitke09} are therefore poor approximations to the
midplane temperatures of our model disks, as are constant-$\alpha$
models in which $T$ decreases with $R$ at the midplane.  Instead,
coupling the simple viscosity prescription in Section
\ref{sec:prescription} with a radiative transfer scheme is preferable
for modeling observables that trace the inner $\sim 70$~AU.

Figure \ref{fig:temperature} shows that the disk surface cools off over
time, as expected for any disk being irradiated by a T-Tauri
star evolving along the Hayashi track. The surface cooling
affects the disk structure as follows:
\begin{enumerate}

\item The overall photosphere height of the disk decreases with time
(Equation \ref{eq:grazing}; Figures \ref{fig:visc30AU},
\ref{fig:dens4AU} and \ref{fig:temperature});

\item The viscosity in the surface layers decreases with time
(Equation \ref{eq:alphascale}; Figure \ref{fig:visc30AU}).

\end{enumerate}
The decrease in surface viscosity with time is due to the
increasing density in the surface layers as the disk cools.

Despite the fact that the disk surface cools with time, the tendency of
the MRI to pile up mass unevenly leads the temperature to increase with
time in certain parts of the disk. The bottom panels of Figure
\ref{fig:temperature} show the temperature increase in the pileup at
1~AU in Model 2. Though modest, the temperature increase does affect the
location of the ice line, which we discuss in the next section.

\subsection{The Ice Line in MRI-Active Disks}
\label{sec:iceline}

Ice forms in protoplanetary disks when the temperature falls below
145-170~K, depending on the water vapor's partial pressure
\citep{PZ04,Lecar06}. Observations of the outer asteroid belt place the
ice line in today's Solar System at 2.7~AU.  Theoretical estimates of
the ice line location in the solar nebula place it a minimum of 0.6~AU
from the sun \citep{davis05, garaud07} and a maximum of 6~AU at the
beginning of the T-Tauri phase, moving inward as the disk evolves
\citep{DR09}. In constant-$\alpha$ disks, the ice line moves inward with
time as the optically thick disk radiates away its accretion energy. At
late times, however, the inner disk loses enough mass to become
optically thin to stellar irradiation, causing the ice line to move
outward with time \citep{garaud07, oka11}.  The more massive a
constant-$\alpha$ disk, the higher its midplane temperature will be.
Taking the \citet{davis05}, \citet{garaud07} and \citet{oka11} disk
models up to a reasonable planet-forming mass of $0.04 M_{\odot}$ or
higher \citep{thommes08} would push their ice lines outside the
terrestrial planet-forming region, more in agreement with the results of
\citet{DR09}.

\begin{figure*}[ht]
\centering
\begin{tabular}{cc}
\includegraphics[trim=1cm 2cm 0cm 14cm, clip=true,
width=0.48\textwidth]{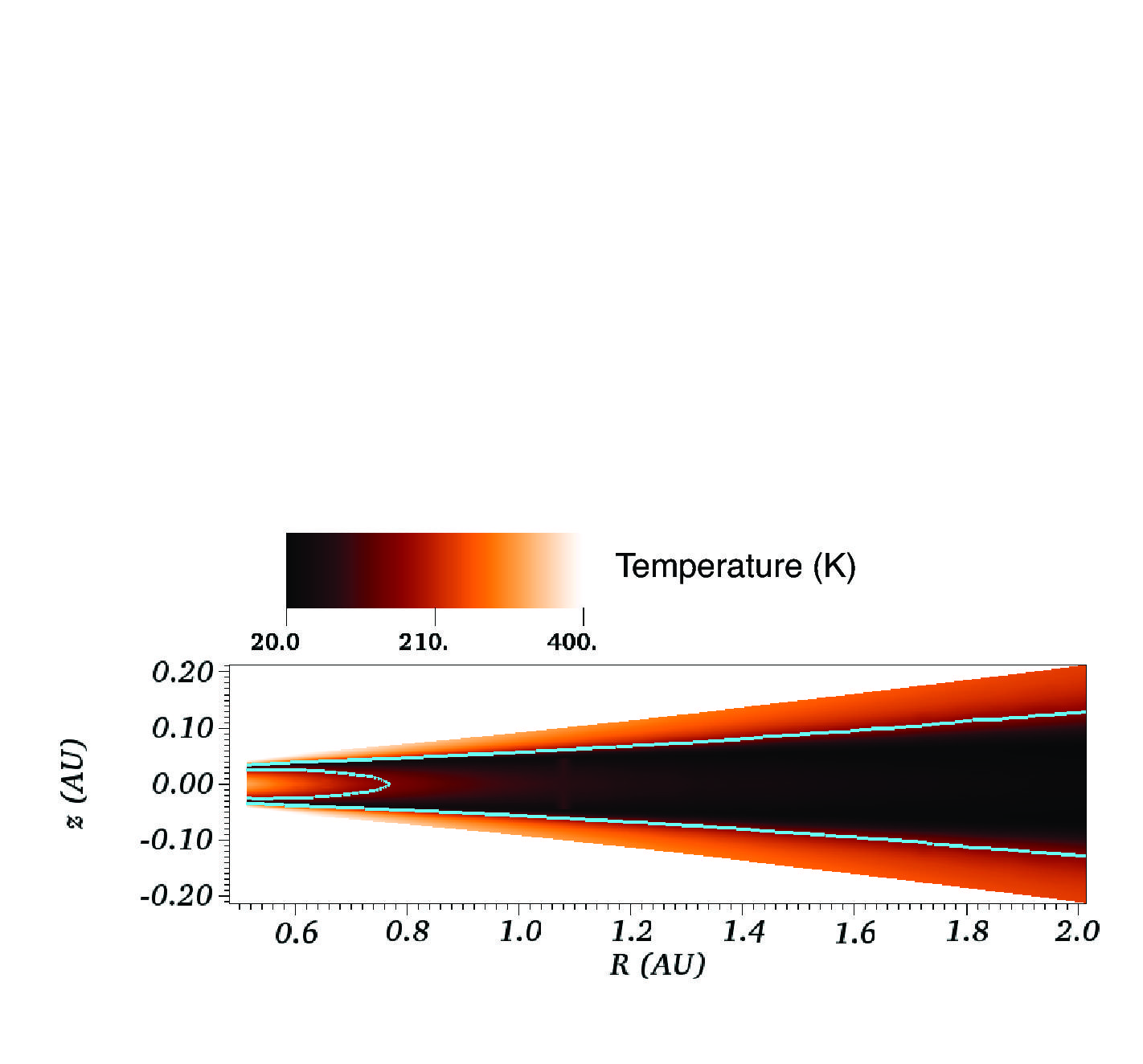}
\includegraphics[trim=1cm 2cm 0cm 14cm, clip=true,
width=0.48\textwidth]{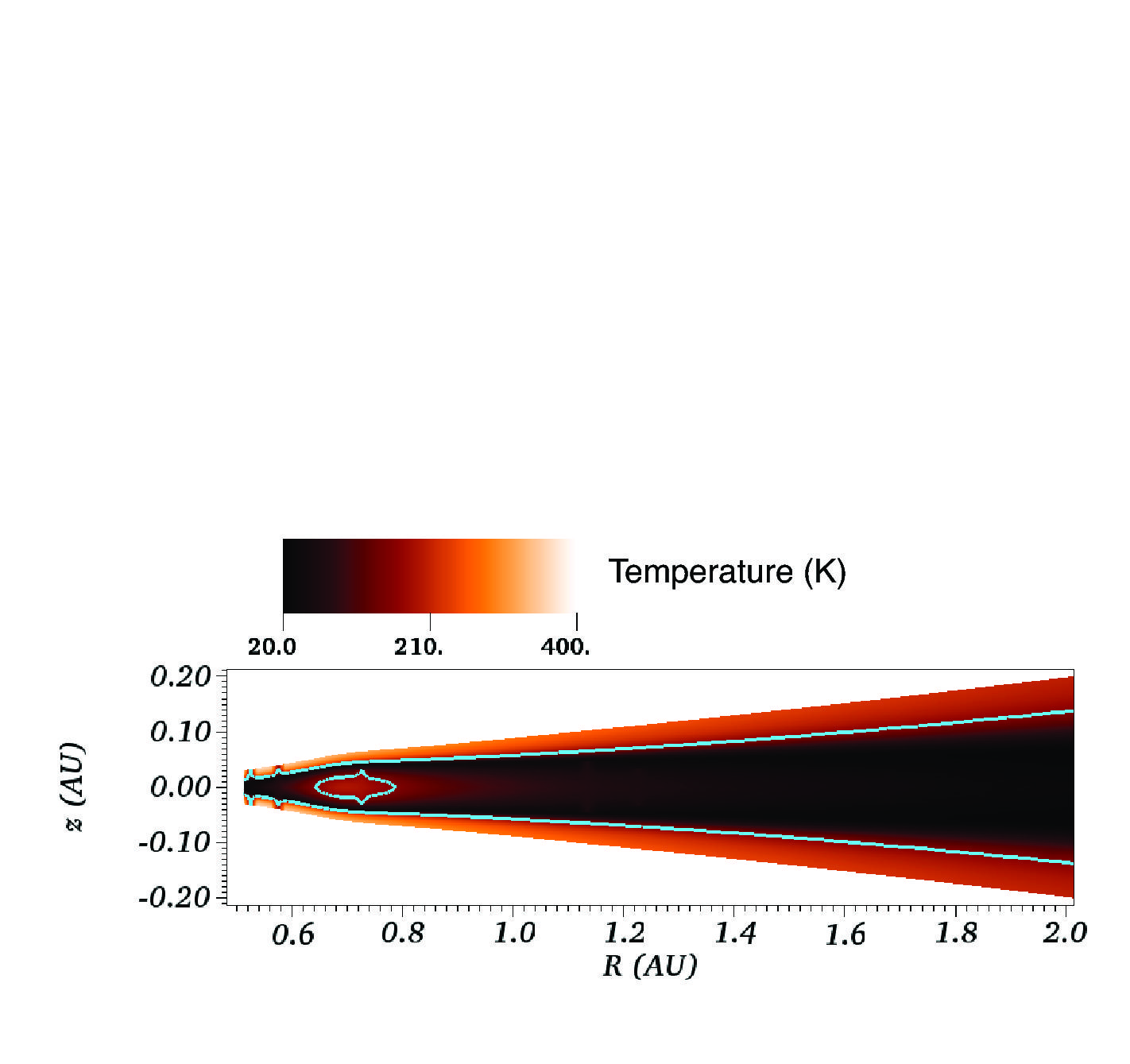} \\
\includegraphics[trim=1cm 2cm 0cm 14cm, clip=true,
width=0.48\textwidth]{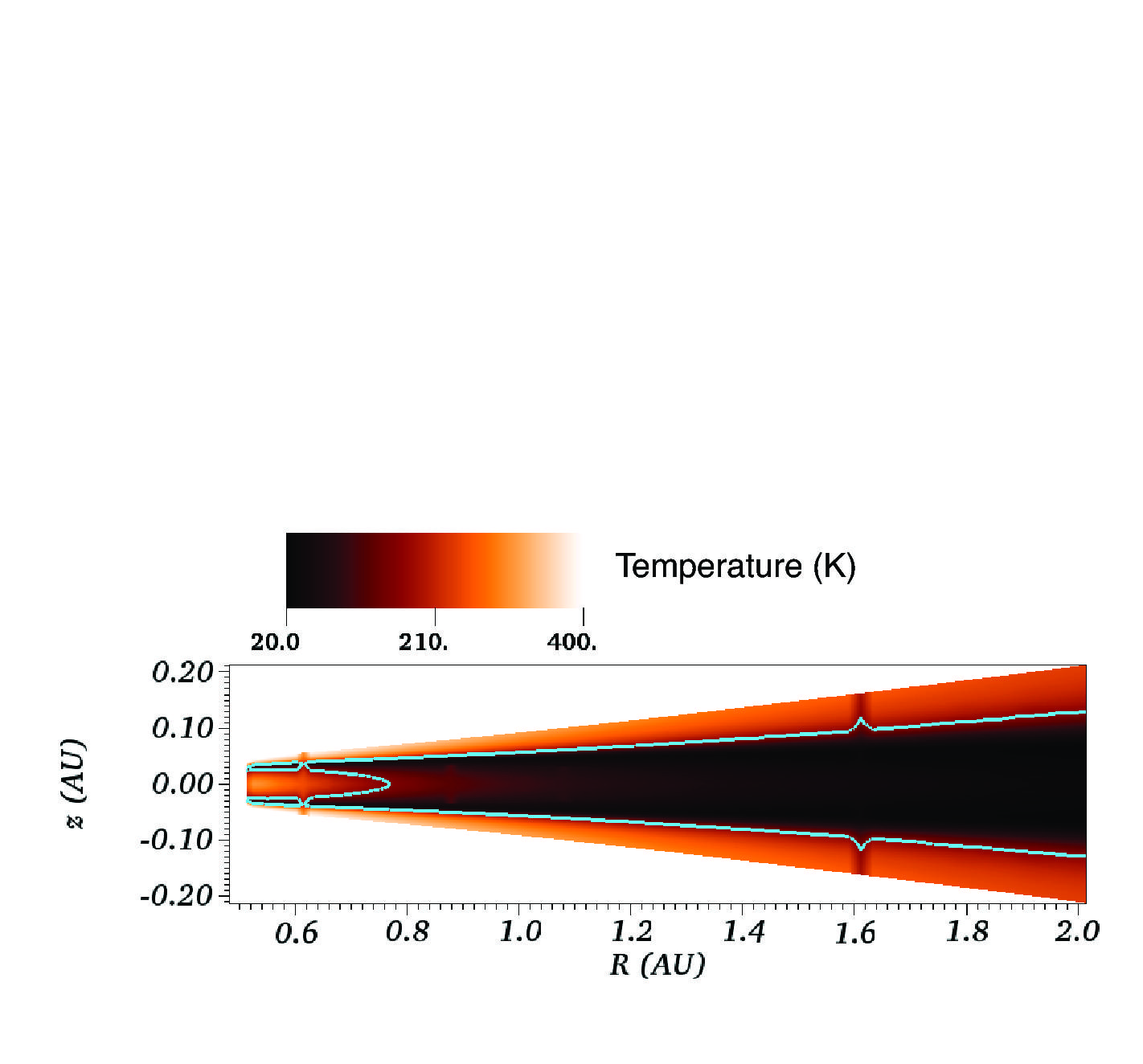}
\includegraphics[trim=1cm 2cm 0cm 14cm, clip=true,
width=0.48\textwidth]{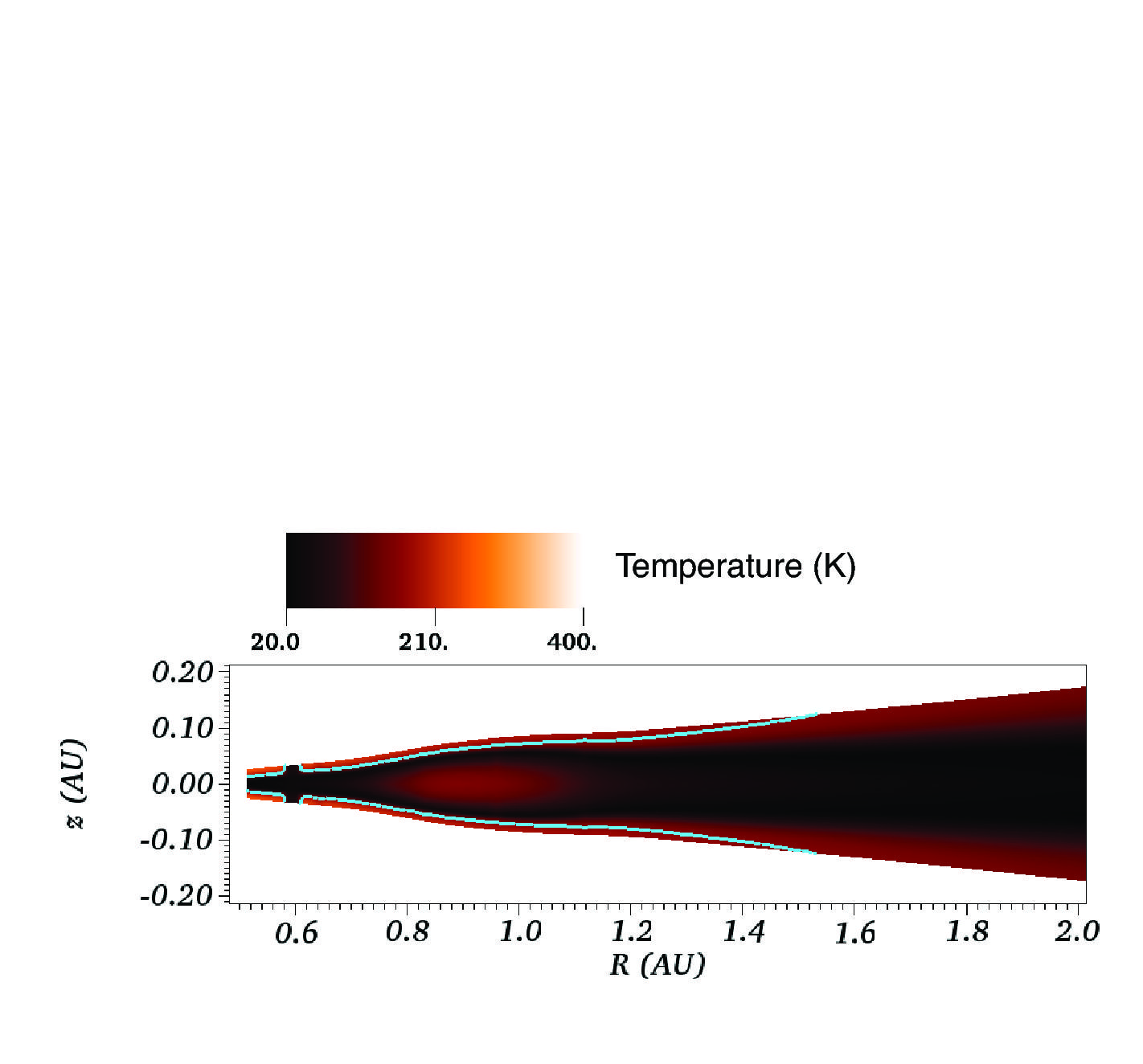} \\
\includegraphics[trim=1cm 2cm 0cm 14cm, clip=true,
width=0.48\textwidth]{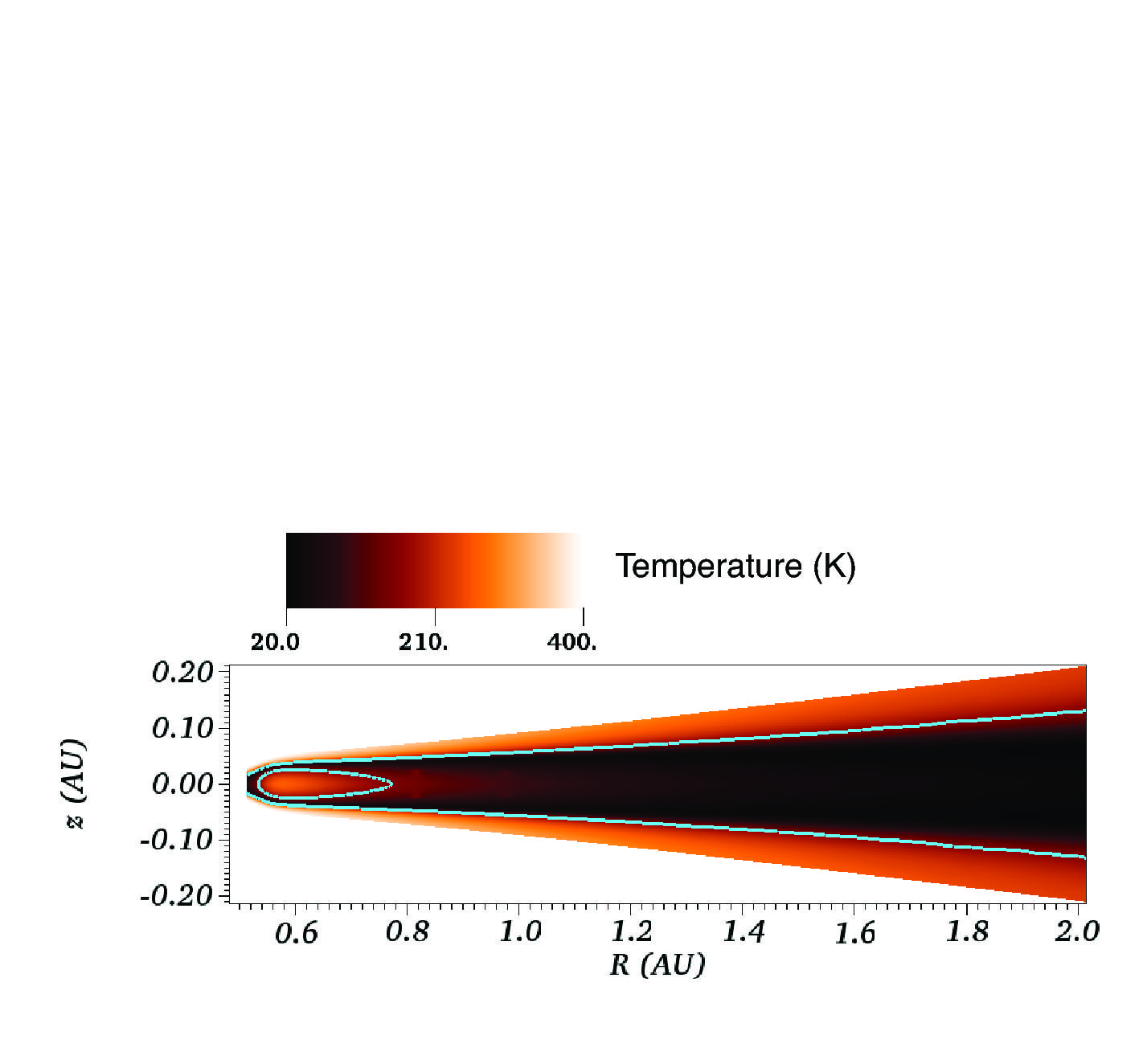}
\includegraphics[trim=1cm 2cm 0cm 14cm, clip=true,
width=0.48\textwidth]{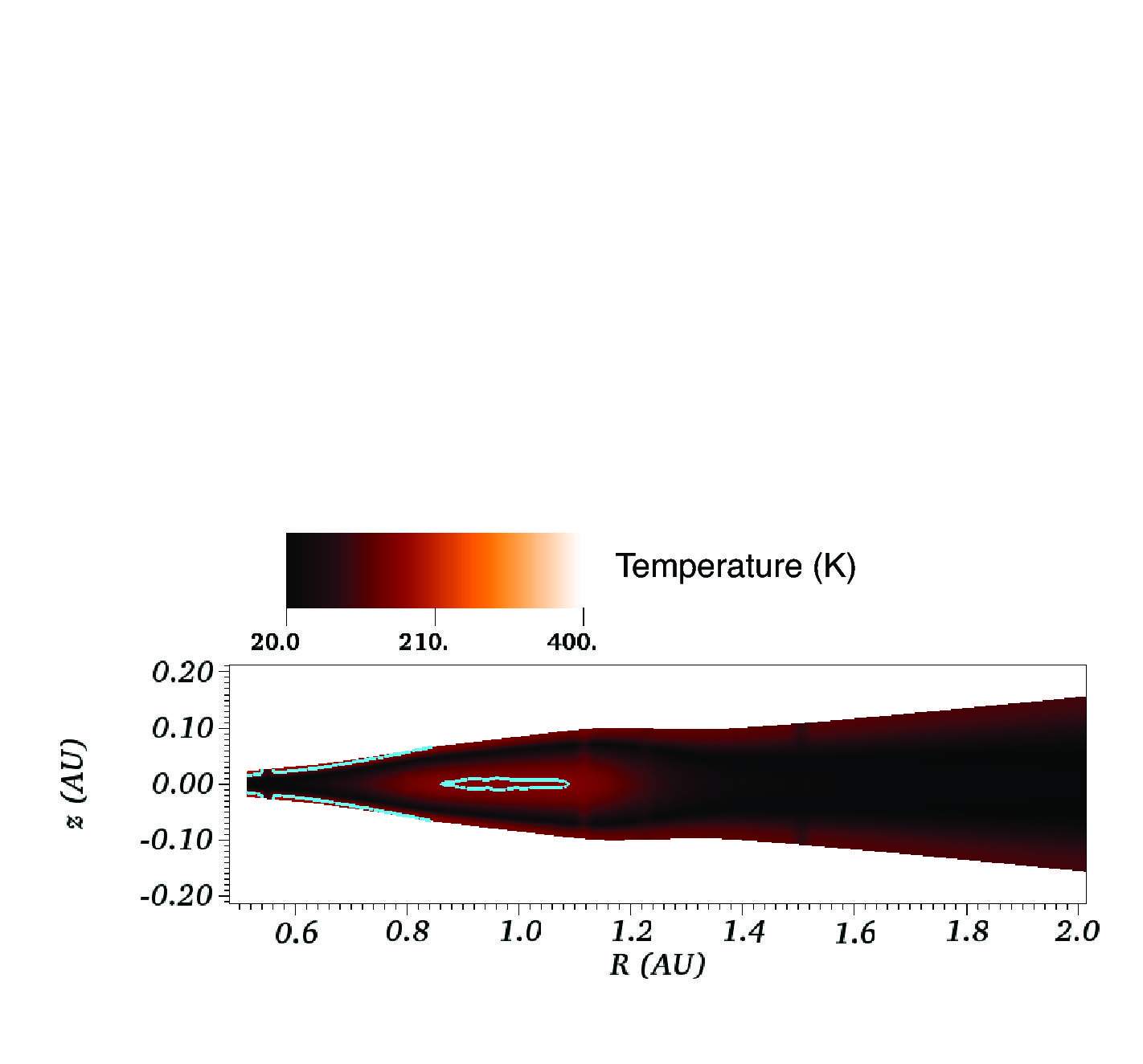}
\end{tabular}
\caption{In the inner 2~AU of Model 2, we see that the disk surface
cooling and midplane heating both affect the structure of the ice line.
Each plot shows $T(R,z)$ with the ice line at 160~K indicated by the
blue contour (temperature units are Kelvins). Simulation times are
100~yr ({\bf top left}), 1000~yr ({\bf middle left}), 10,000~yr ({\bf
bottom left}), 100,000~yr ({\bf top right}), 1~Myr ({\bf middle right}) and
3~Myr ({\bf bottom right}). Recall that the star age is already 0.1~Myr
at the beginning of the simulation.}
\label{fig:iceline}
\end{figure*}

In an MRI-active disk, the overall thermal evolution is
determined by the dimming of the parent star, which causes viscosity in
the active surface layers to decrease with time (see Section
\ref{sec:heatcool}).  Figure \ref{fig:iceline} shows the 2-D structure
of the ice line in Model 2 at timepoints 100~yr, 1000~yr, 10,000~yr,
100,000~yr, 1~Myr and 3~Myr.  At early times, we recover the ``two-branch''
structure of the ice line seen by \citet{davis05}, in which a nearly
horizontal ice line divides the hot surface from the cool interior and a
midplane ice line separates the warm midplane near the star from the
cool midplane far away from the star. After 10,000~yr of evolution, the
inner edge of the disk at 0.5~AU cools enough for ice to freeze at the
midplane, causing a ``pinch-off'' in the midplane branch of the ice line
that leaves an H$_2$O gas bubble at 0.6~AU. This pinch-off is probably a
boundary effect: since mass flows freely from our inner boundary at
0.5~AU onto the star, the disk near our inner boundary loses mass and
cools quickly. 

By 1~Myr, mass loss from the inner edge of our grid combined with less
activity in the surface layers cools the disk enough to push the
midplane ice line inside 0.5~AU, the inner boundary of our computation.
Here our results are still consistent with the low-mass models of
\citet{davis05}, \citet{garaud07} and \citet{oka11}, who all show the
ice line moving inside 1~AU for accretion rates $\dot{M} \sim 10^{-10}
M_{\odot}$~yr$^{-1}$. An important difference between our models and
those of \citet{davis05}, \citet{garaud07} and \citet{oka11} is that the
accretion rate does not have to drop extremely low for the terrestrial
planet-forming region to cool enough to freeze ice: both Model 1 and
Model 2 have active layers that drive $\dot{M} \sim 10^{-9}
M_{\odot}$~yr$^{-1}$ through the disk surface. Our model disks 
predict colder midplanes in the inner 10 AU than those of
\citet{terquem08} due to the lower minimum $\alpha$ assumed here
($\alpha_{min} = 10^{-5}$ for our work vs.\ $\alpha_D = 10^{-3}$ or
$\alpha_D = 10^{-4}$ in Figure 3 of \citet{terquem08}, where
$\alpha_D$ the value of $\alpha$ in the dead zone). Unlike \citet{terquem08},
our disk has a colder midplane than surface because we include the
effects of stellar heating.

The MRI can also create local regions that heat up with time not because
the disk becomes optically thin, but because mass piles up in the dead
zone.  Between 1~Myr and 3~Myr of evolution, the pileup at 1~AU of
Model~2 heats up enough for the midplane ice line to reappear---a real
physical effect. Model~1, which does not show any such pileup, follows a
similar evolutionary path as Model~2 up to 1~Myr. In an MRI-active disk,
water ice in the terrestrial planet-forming region may be transient.
Understanding which parts of the inner disk have water ice available for
planet formation requires a careful comparison of the planetesimal
growth timescale, the star cooling timescale and the growth timescale of
any pileups deposited in the dead zone. In Model~2, the pileup grows and
the star dims on a timescale similar to the disk lifetime. We will
examine high-mass MRI-active disks, in which dead-zone pileups can grow
much more quickly, in a forthcoming paper.

\section{Model Limitations}
\label{section:limits}

Although our model incorporates much of the physics of disk evolution
during the T-Tauri phases of stars, our long-timescale computation
requires a number of simplifications. Our disk accretion model suffers
from the following limitations, which may affect our conclusions:

\begin{enumerate}
\item We model our disks as 1+1d instead of 3d. We assume radial symmetry,
and that the vertical structure is not coupled to the radial mass transport.
This ignores the possibility of accretion driven by non-axisymmetric 
instabilities not represented by our parameterization of stress from MHD
turbulence, such as the Rossby wave instability \citep{hawley87, li01,
meheut12}. We may then underestimate accretion rates in some
regions of the disk.

\item We do not model the effects of gravitational instability on
momentum transport, as is done by \citet{martin12b}, but instead
limit our simulations to those disks which are gravitationally stable for
their entire lifetime. To test whether a disk is
gravitationally stable to axisymmetric perturbations, we calculate the
Toomre Q parameter at every radial grid point. Stability requires that
\begin{equation}
\label{eq:toomreQ}
Q \equiv \frac{c_{s} \Omega}{\pi G \Sigma} > Q_{crit} \approx 1.
\end{equation}
Note that for disks more massive than those we have simulated, the disk
does eventually become gravitationally unstable to axisymmetric 
perturbations within the dead zone (see Section \ref{sec:mass_flow}).

\item We do not model the effects of Hall diffusivity, which is the least
well understood magnetic diffusion regime. However, it is
unlikely to alter either the conditions required for MRI or the strength
of the turbulence where Ohmic dissipation is also present \citep{SS02}.

\item We make several assumptions about the magnetic field strength and
presence of grains. The activity of MRI driven turbulence is very
sensitive to these parameters. First, we deviate from the standard
interstellar gas/small grain mass ratio of 100 and instead assume a
ratio of 1000. This is equivalent to assuming that 90\% of the grain
mass is either in grains larger than 1$\mu$m
\citep{oliveira10,scaife12}, which do not significantly affect electron
density, or has settled below the dead zone. \citet{Mohanty13} find that
grain depletion through growth or settling is required to account for
the observed accretion rates of low mass protostars. Without the high
gas/small grain mass ratio, the inner Ohmic dead zone and the upper
Ambipolar dead zone overlap, producing a passive thermal structure for
much of the radial extent of the disk (see Section
\ref{sec:turbulent_structure}).

We also assume a vertically constant magnetic pressure with a midplane plasma 
$\beta$ of 1000. This is at the upper end of the range for midplane plasma
$\beta$ found by \citet{FN06} in global MHD simulations with saturated
turbulence. Without a sufficiently weak magnetic field, the
ambipolar diffusivity can be large enough that there is once again an
overlap in the dead zones and an at least partially passive disk.

\item We assume a zero stress boundary condition at the inner radius of
our disks (see Section \ref{subsec:methods}). This creates a non-realistic
boundary effect in which mass is rapidly depleted from the inner annuli. One
result of this can be seen in Figure \ref{fig:iceline}, as the midplane
within 0.6 AU becomes cold due to the loss of surface density.

\end{enumerate}

Due to computational convergence difficulties in our model, at any particular
timestep there are a few annuli with vertical structures which do not meet one
or more of our midplane boundary conditions (Equations \ref{eq:midplane_flux},
\ref{eq:dens_converge}, and \ref{eq:constant_beta}), due to the discontinuous
nature of the MRI and opacity. We do not
consider this to be a limitation on our simulations, as the unsolved annuli's
contribution to the viscosity profile is smoothed with a Gaussian filter before
being used to update the surface density profile (Equation \ref{eq:diffusion}).
Although there are always a few ``bad'' annuli present, at any particular radius
the lack of convergence for the vertical structure persists for only a few
timesteps. The unconverged annuli can be seen as occasional incongruous vertical bars in our contour
plots (Figures \ref{fig:visc30AU}, \ref{fig:vstructzoom}, \ref{fig:dens4AU},
\ref{fig:temperature}, and \ref{fig:iceline}). 

\section{Conclusions}
\label{sec:conclusions}

In the Introduction, we asked five questions about the structure
and evolution of MRI-active disks. Here we summarize our
findings and answer each question:

\begin{enumerate}

\item How do the relative sizes of the dead zone and active
layers change over time?

The radial size of the dead zone is almost constant in time, while the
vertical height of the dead zone shrinks over time.  What was surprising
about our results was not the evolution of the dead zone, but the
complexity of the disk structure. Between 1.2~AU and 1.7~AU, Model 2 has
a five-layer structure throughout most of its evolution: inactive
corona, active layer, dead slice, active layer, dead midplane (see
Section \ref{sec:turbulent_structure} and Figures \ref{fig:vstructline}
and \ref{fig:vstructzoom}). Note, however, that detailed 3-D simulations
would likely show no dead slice since the MRI wavelength in the active
layers bracketing the dead slice is of order the dead slice thickness.
The lower-mass Model 1 has at most three layers: inactive corona, active
layer, dead midplane. Throughout this work, we have seen that increasing
disk mass leads to increasing complexity in the disk structure and
accretion flow.

Finally, the radial size of the dead zone was somewhat higher than
predicted in previous work: 16~AU for Model 1 and 21~AU for Model~2.
Previous papers reporting a $\sim 5$~AU dead zone used the MMSN
\citep[e.g.][]{matsumura03, salmeron08, flaig12}, but the modestly
higher surface densities of our disks expanded the dead zone. There is
some evidence that the solar nebula had a large dead zone consistent
with our findings---the giant planets have an atmospheric composition
gradient that, if primordial, would have diffused on million-year
timescales if not protected by a dead zone \citep{nelson10}. One caveat,
though, is that we have assumed that the stellar X-ray flux is constant
in time, as is the cosmic ray flux. A decreasing stellar X-ray flux,
which ionizes mainly the inner $\sim 3$~AU of the disk surface, might
erase the dead slice over time, while changing the cosmic ray flux as
the ambient molecular cloud disperses would certainly change the radial
extent of the dead zone.

\item How does $\dot{M}$ vary with radius and time?

Throughout the disk evolution, $|\dot{M}|$ is highest at the inner
and outer boundaries of the dead zone. The lowest $|\dot{M}|$ in the
inner disk, where gas flows toward the star, is in the middle of the
dead zone. $|\dot{M}|$ is about 50\% higher for Model~2 than Model~1,
suggesting that higher-mass disks support higher MRI-driven accretion
rates---though the increase in $\dot{M}$ with $M_{disk}$ is modest.
$|\dot{M}(t)|$ decreases extremely slowly: though the depth of the
active layer does not change with time (Figure \ref{fig:activedepth}),
the viscosity in the active layer drops modestly as the star cools.

As required by conservation of angular momentum, both model
disks expand as they evolve, creating a turn-around in the
accretion flow. The turn-around radius is almost entirely
determined by $R_{out}$ at $t = 0$ and moves inward as the disk
evolves. Here, with $R_{out} = 70$~AU at $t = 0$, the
turn-around radius eventually reaches 40~AU after 1~Myr of
evolution. Note that the turn-around radius moves steadily
inward in both models and does not converge toward a particular
location (Figure \ref{fig:mdot}). One expects the turn-around
radius to move steadily inward because mass must continually
join the outward ``decretion'' flow in order to keep
transporting angular momentum outward.

Our models predict $\dot{M} \sim 10^{-9} M_{\odot}$~yr$^{-1}$ in the
planet-forming region of the disk, but the exact value of $\dot{M}$
depends on many free parameters such as grain size, gas/small grain mass
ratio and magnetic field strength. We have not attempted an exhaustive
parameter study of $\dot{M}$ as a function of all variables. We merely
note that for a gas/small grain mass ratio of 100 and a grain size of
$0.1 \mu$m, all MRI activity in the disk was suppressed. Likewise,
decreasing plasma $\beta$ at the midplane from 1000 to 100 suppressed
MRI turbulence, though not as severely as small grains.



\item How can disk modelers parameterize heating in the active
layers and dead zone without resorting to a 3-D MHD simulation?

The value of non-evolving, ``snapshot'' disk models is inarguable,
particularly for modeling observables. Section \ref{sec:prescription}
presents a simple modification of the standard, constant-$\alpha$
irradiated disk that approximates the thermal structure of the disk
where the dead zone is present. Simply set the surface density of the
active layer to 10~g~cm$^{-2}$ and use $\alpha \approx 0.01$ (see
Figures \ref{fig:vstructline}, \ref{fig:vstructzoom} and
\ref{fig:activedepth}).  For the dead zone, use $\alpha \approx
10^{-5}$. For numerical models, we recommend smoothing the transition
between the dead zone and the active layer. While the depth of the
active layer does vary across the dead zone, its variation is at most a
factor of two in a given disk. The fact that the active layer depth is
almost constant in time for both Model~1 and Model~2 makes our simple
prescription applicable to any stage of T-Tauri disk evolution, provided
that stellar irradiation is included.

\item Does the disk midplane heat up or cool off with time?

In both models, the midplane temperature varies little with time, but
the disk surface cools as the star evolves down the Hayashi track.
Despite the modest masses of our model disks, the optical depth of both
to stellar irradiation is enough to thermally decouple the surface from
the midplane. Most of the dead zone is so lacking in energy generation
that it falls to the assumed ambient temperature of surrounding molecular
cloud material, 20~K in these models. Since the radial extent of the
dead zone changes little with time, the disk midplane temperature
remains static except at $R \ga 60$~AU, where the disk thins enough
over time to become optically thin to stellar irradiation (Figure
\ref{fig:temperature}).

In the midplane, there are two possible locations where the temperature
is not static but increases with time (Figure \ref{fig:temperature}).
The first is at the outer edge of the dead zone. A slight decrease in
surface density with time pushes the dead zone boundary modestly inward,
allowing material at the edge of the dead zone to become turbulent and
heat up. The other location of increasing temperature with time is the
pileup at 1~AU of Model 2.  Lower-mass Model 1 does not develop any such
pileups on million-year timescales. To the extent that MRI-driven
accretion deposits piles of material in the dead zone, the disk midplane
may heat up.

Note, however, that the thermal properties of the disk depend on
the ionizing radiation it receives. Near the star, the disk
structure would evolve if the X-ray flux were to change with
time. In future work, it would be interesting to let $L_X$ scale
with bolometric luminosity on the Hayashi track. A spike in
cosmic ray ionization from a nearby supernova would affect the
global disk structure \citep{fatuzzo06}. 

\item Where is the ice line in an MRI-active disk and how does
its location change over time?

Due to the paucity of energy generation in the dead zone, the midplane
ice line falls somewhere inside the terrestrial planet-forming region.
In our models, the midplane ice line actually moves off the inner edge
of our grid at 0.5~AU after $10^5$ years of disk evolution, reappearing
in Model 2 after the pileup at 1~AU reheats the midplane. Despite the
different physics used in computing the ice line location, our results
roughly agree with those of \citet{davis05}, \citet{garaud07} and
\citet{oka11} in that the midplane ice line is inside 1~AU for most of
the disk's evolution. Determining whether and when ice is available for
terrestrial planet formation requires carefully comparing the disk
cooling timescale, planetesimal growth timescale and the timescale on
which MRI-deposited pileups grow. An important difference between our
model and standard, constant-$\alpha$ disk models is that the accretion
rate does not have to drop extremely low to move the ice line inside
1~AU: throughout their evolution, our disks have $\dot{M} \ga 10^{-9}
M_{\odot}$~yr$^{-1}$ moving through the active layers that sandwich the
icy inner disk.

\end{enumerate}

Here we have presented the first analysis of the structure and evolution
of an entire MRI-active disk on million-year timescales. While we have
chosen to focus on low-mass disks in this work in order to compare with
previous studies, we will expand our analysis to include high-mass,
planet-forming disks in a forthcoming paper. Already, at $M_{disk} =
0.03 M_{\odot}$---just short of the minimum $0.04 M_{\odot}$ required
for giant planet formation \citep{thommes08}---Model~2 exhibits some new
features such as the split active layer and the re-heating midplane.

Funding for this work was provided by NASA through grant NNX10AH28G to
S.D.R. and N.J.T., and by University of Texas through a startup grant to
S.D.R. Computing and visualization support were provided by the Texas
Advanced Computing Center, which is funded by the National Science
Foundation. N.J.T.\ carried out his work at the Jet Propulsion
Laboratory, California Institute of Technology, under a contract with
NASA.

\begin{deluxetable}{ccll}
\tabletypesize{\tiny}
\tablecaption{Parameters in protostellar disk model
\label{table:diskpars}}
\tablehead{
\colhead{Name} & \colhead{Value} & \colhead{Description} &
\colhead{Reference}
}
\startdata

\multicolumn{4}{l}{Ionization parameters} \\
$\zeta_{CR}^{surf}$ & $10^{-17}$~s$^{-1}$ & Cosmic ray
ionization rate at disk surface & \citet{UN81} \\
$\lambda_{CR}$ & 96~g~cm$^{-2}$ & Cosmic ray penetration depth &
\citet{UN81} \\
$L_X$ & $2 \times 10^{30}$~erg~s$^{-1}$ & Stellar X-ray
luminosity & \citet{garmire00} \\
$\zeta_1$ & $6.0 \times 10^{-12}$~s$^{-1}$ & Ionization
rate coefficient for absorbed X-rays & \citet{BG09} \\
$\zeta_2$ & $1.0 \times 10^{-15}$~s$^{-1}$ & Ionization rate
coefficient for scattered X-rays & \citet{BG09} \\
$\lambda_1$ & $2.5 \times 10^{-3}$~g~cm$^{-2}$ & Penetration
depth of absorbed X-rays & \citet{BG09} \\
$\lambda_2$ & 1.2~g~cm$^{-2}$ & Penetration depth of scattered
X-rays & \citet{BG09} \\
$p_1$ & 0.4 & Exponent of absorbed X-ray attenuation & \citet{BG09}
\\
$p_2$ & 0.65 & Exponent of scattered X-ray attenuation &
\citet{BG09} \\

\multicolumn{4}{l}{Ambient medium} \\
$T_{amb}$ & 20 K & Ambient temperature set by remnant molecular
cloud & \citet{peretto10} \\

\multicolumn{4}{l}{Dust grains} \\
$\rho_{gr}$ & 3.0~g~cm$^{-3}$ & Internal grain density &
standard \\
$G/S$ & 1000 & Gas/small grain mass ratio & augmented from
standard 100 to approximate grain growth \\
$a$ & 1 $\mu$m & Grain size & \citet{oliveira10} \\

\multicolumn{4}{l}{Gas composition} \\
$N_{\mathrm Mg}$ & $10^{-4} N_{{\mathrm Mg},\odot}$ & Magnesium
abundance in disk gas & \citet{TS08} \\
$\mu$ & 2.33~g~mol$^{-1}$ & Mean molar weight & standard \\

\multicolumn{4}{l}{Maxwell stresses} \\
$\alpha_{min}$ & $10^{-5}$ & Minimum stress in dead
zone from large-scale fields & \citet{TS08} \\
\enddata
\end{deluxetable}

\clearpage


\end{document}